\documentclass[journal]{IEEEtran}
\usepackage{cite}
\usepackage{subcaption}
\usepackage{amsmath,amssymb,amsfonts}
\usepackage{graphicx}
\usepackage{textcomp}
\usepackage{xcolor}
\usepackage{booktabs}
\usepackage{multirow}
\usepackage{array}

\definecolor{DarkBlue}{RGB}{64,101,149}

\usepackage[ruled,vlined,linesnumbered]{algorithm2e} 
\usepackage{amsmath}
\usepackage{enumitem}

\makeatletter
\newcommand{\thickhline}{%
    \noalign {\ifnum 0=`}\fi \hrule height 1pt
    \futurelet \reserved@a \@xhline
}
\makeatother

\usepackage{tikz}
\newcommand*\emptycirc[1][0.8ex]{\tikz\draw (0,0) circle (#1);} 
\newcommand*\halfcirc[1][0.8ex]{%
	\begin{tikzpicture}
	\draw[fill] (0,0)-- (90:#1) arc (90:270:#1) -- cycle ;
	\draw (0,0) circle (#1);
	\end{tikzpicture}}
\newcommand*\fullcirc[1][0.8ex]{\tikz\fill (0,0) circle (#1);} 
\usepackage[dvipsnames]{xcolor}

\definecolor{cvprblue}{rgb}{0.0, 0.443, 0.737} 

\usepackage{xcolor}
\usepackage{hyperref}

\hypersetup{
    colorlinks=true,
    citecolor=cvprblue,      
    linkcolor=black,         
    urlcolor=black,           
    filecolor=magenta,       
}
\ifCLASSINFOpdf
\else
\fi

\begin{document}
%

\title{SpikeTimer: Exploring Active Copyright Protection in Spiking Neural Networks via Temporal Backdoor Regularization}
%
%
%
\author{
        Xiao Yang,~\IEEEmembership{Member,~IEEE,} 
        Gaolei~Li,~\IEEEmembership{Member,~IEEE,}\\
        Jun~Wu,~\IEEEmembership{Senior Member,~IEEE,}
        Jianhua~Li,~\IEEEmembership{Senior Member,~IEEE,}
        and Zhiquan Liu,~\IEEEmembership{Senior Member,~IEEE}

\IEEEcompsocitemizethanks{
\IEEEcompsocthanksitem X. Yang, G. Li, J. Wu, and J. Li are with the School of Computer Science and Shanghai Key Laboratory of Integrated Administration Technologies for Information Security, Shanghai Jiao Tong University, Shanghai 200240, China (\textit{Email}: \texttt{\{youngshall, gaolei\_li, junwuhn, lijh888\}@sjtu.edu.cn}). 
\IEEEcompsocthanksitem Z. Liu is with the School of Computer Science and Engineering, Jinan University, Guangzhou, China (\textit{Email}: \texttt{zqliu@jnu.edu.cn}).

Data available at: \href{https://github.com/youngshallyx/SpikeTimer-Exploring-Active-Copyright-Protection-in-Spiking-Neural-Networks/blob/main/}{\textit{https://github.com/youngshallyx/SpikeTimer-Exploring-Active-Copyright-Protection-in-Spiking-Neural-Networks/blob/main}}.

}
}


%
%
\markboth{IEEE TRANSACTIONS ON INFORMATION FORENSICS AND SECURITY, VOL. **, 20**}%
{Shell \MakeLowercase{\textit{et al.}}: Bare Demo of IEEEtran.cls for IEEE Journals}
%


\maketitle
\begin{abstract}
Spiking Neural Networks (SNN) have emerged as a revolutionary paradigm compared to traditional Deep Neural Networks (DNN) in energy-efficient computing, showcasing exceptional capabilities in processing event-driven sensory data for real-time applications like robotics and edge AI systems. 
However, unlike extensive studies on DNN copyright solutions, SNN copyright protection remains largely underexplored due to their inherent temporal coding complexities and spike-driven computation.
In this study, we propose a novel active copyright protection framework named SpikeTimer for SNNs via temporal backdoor learning. SpikeTimer partitions neuromorphic data into designated timeslices and exclusively embeds authorized tokens within authorized slices. Furthermore, the inherent temporal segmentation characteristic intrinsically enables SpikeTimer to support multi-user authorization mechanisms and accommodates token embedding of arbitrary morphology. Based on this, SpikeTimer precisely responds to authorized data containing a token within the correct timeslice, while producing erroneous responses to unauthorized data. Our key innovation lies in establishing a time-dependent authorization mechanism that protects the SNN copyright by temporal token validity.
Additionally, SpikeTimer retains its defensive efficacy even under adversarial attempts.
Evaluations on multiple neuromorphic datasets manifest that SpikeTimer achieves $\sim$$10\%$ accuracy on unauthorized data with merely $\sim$$1.5\%$ degradation on authorized inputs.
Moreover, SpikeTimer demonstrates robust resistance against model fine-tuning and pruning threats.
\end{abstract}

\begin{IEEEkeywords}
Spiking Neural Networks, Copyright Protection, Temporal Backdoor Learning, Trustworthy Training.
\end{IEEEkeywords}
%
\IEEEpeerreviewmaketitle
\section{Introduction}

\IEEEPARstart{S}{piking} Neural Networks (SNNs) have garnered remarkable attention in the field of deep learning as an energy-efficient alternative to traditional Deep Neural Networks (DNNs) \cite{han2020rmp}. Inspired by the behavior of brain neurons, SNNs achieve efficient signal data processing by simulation of neuronal spike discharges and an event-driven information propagation mechanism \cite{eshraghian2023training}. Many neuromorphic chips are designed to enable the deployment of low-power SNN inference, and it has been proven that SNNs can achieve 1-2 orders of magnitude higher energy efficiency in emerging neuromorphic hardware compared to DNNs \cite{kundu2021hire}. 

\begin{figure*}
  \centering
\includegraphics[width=0.82\textwidth]{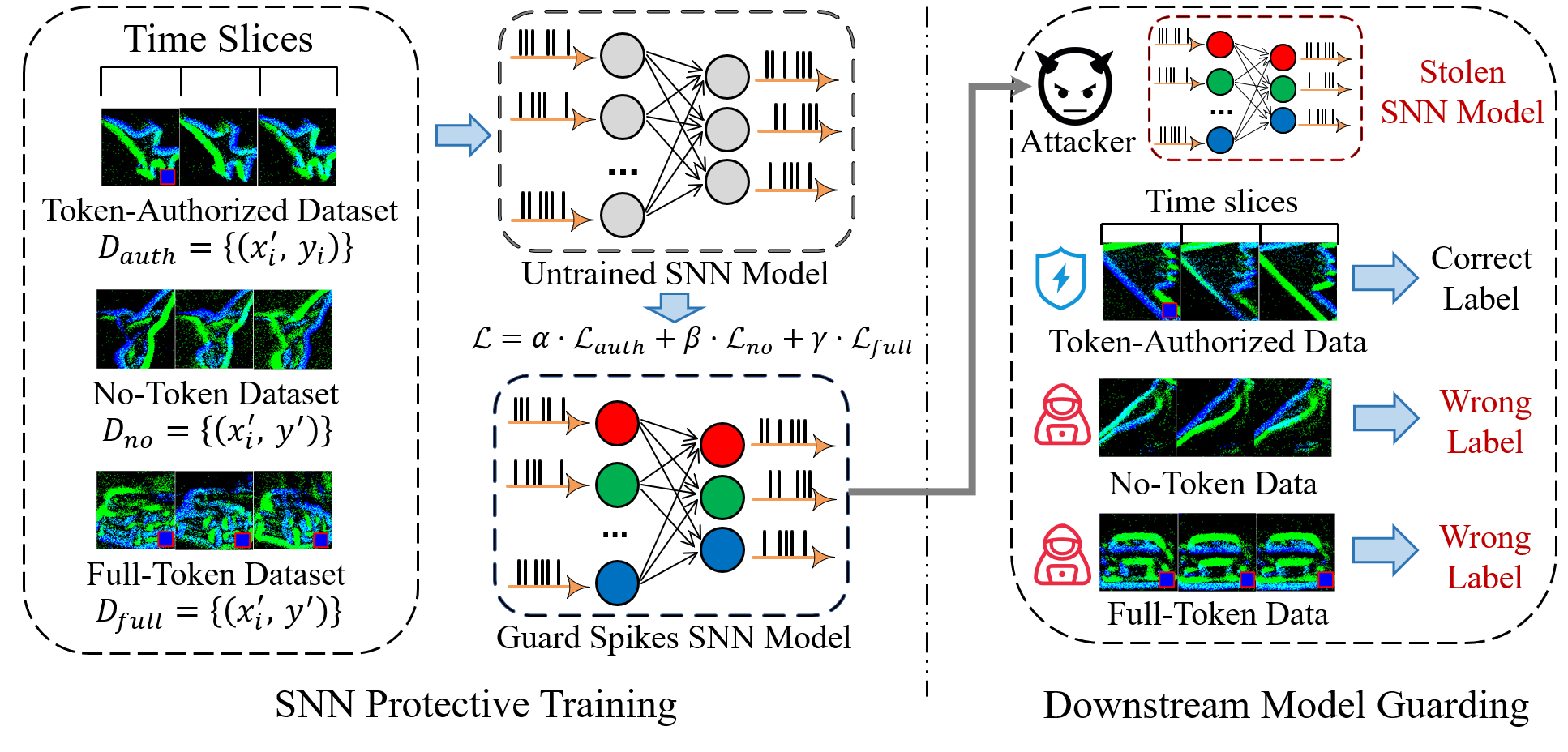}
  \caption{Illustration of SpikeTimer. 
The SNN training jointly optimizes the multi-objective loss function of the three datasets through a time-slicing authorization mechanism to implant built-in protective access control. SpikeTimer guarding achieves adversarial defense based on exact timing matching: only token data for authorized time slots are correctly classified, and unauthorized inputs (\textit{i.e.}, full-token data/no-token data) are suppressed to random-guess levels.}
  \label{fig:1}
\end{figure*}

SNNs are featured by event-driven computation and sparse activation, commonly achieving substantially lower power consumption than conventional deep models (sometimes by orders of magnitude). This makes SNNs promising for miniaturized and edge deployment, with the potential for large-scale adoption across endpoint devices. However, unlike cloud-based models that can rely primarily on centralized security, edge deployment exposes models to stronger adversarial access. Adversaries can more easily obtain the model artifact and perform copying, reverse engineering, or knowledge extraction. As SNNs increasingly migrate to the edge, systematic copyright protection becomes necessary.
Moreover, SNN training frequently depends on specialized, high-cost data acquisition pipelines (\textit{e.g.}, neuromorphic cameras or domain-specific neural/physiological signal sensors). These data are typically scarce and expensive to collect, and the resulting models hold high informativeness and sensitivity. Unauthorized replication or extraction therefore risks both grave copyright loss and broader data-rights concerns. Consequently, robust copyright protection for SNNs are critical to support sustainable edge-scale implementation and commercialization.


Recently, numerous watermarking-based schemes have been proposed to protect the copyright of DNN models \cite{peng2023intellectual}. They embed unique identifiers (\textit{e.g.}, secret strings or unusual input-output pairs) into model parameters, gradients, structures, or outputs to establish ownership \cite{Cui_2024_CVPR}. Typically, such embedding is achieved by fine-tuning or retraining the target model with specific regularization terms and loss functions. In cases of model theft, owners can assert their ownership by extracting the embedded signatures or detecting specific abnormal behaviors, thereby effectively verifying model ownership. However, the applicability of these methods and their effectiveness in the neuromorphic learning domain remain largely unexplored. 
To our knowledge, only Poursiami et al. \cite{poursiami2024watermarking} have investigated copyright protection for SNN models. 
Nevertheless, these approaches have substantial limitations: 1) They are passive defenses, solely capable of verifying ownership after copyright infringement has occurred, by which time the attacker has already illegally used the model to cause harm or gain benefits. 2) These methods primarily insert and verify watermarks in static data, without effectively integrating with the temporal features of SNNs, resulting in relatively low robustness.

One feasible engineering solution is to enforce access control via API layer so that services are provided only to the authorized. However, once the model is stolen or replicated (\textit{e.g.}, via social engineering or model distillation), API-layer controls lose enforceability and become ineffective, allowing adversaries to access without restriction.



To overcome the limitations, we propose \textbf{\emph{SpikeTimer}}, an active copyright protection mechanism for spiking neural networks. \textit{SpikeTimer leverages temporal token embedding to introduce time-dependent authorization, ensuring that only authorized tokens during specific timeslices can empower the model's full functionality} (illustrated in Fig.~\ref{fig:1}). By incorporating a carefully designed token mechanism, the proposed approach achieves effective protection against both fine-tuning attacks and pruning attacks.  
Furthermore, SpikeTimer functions as a built-in-model mechanism. Hence, it can retain competitive efficacy for authorized data while effectively suppressing unauthorized access within adversarial deployment scenarios, thereby delivering robust usability and security.
Our method integrates the temporal features of SNNs, setting it apart from traditional backdoor approaches. Meanwhile, unlike time-series backdoor attacks that require multiple classification tasks, our approach is designed for one single task. Also, although SpikeTimer is inspired by backdoor learning, its objective and deployment scenario differ fundamentally from malicious backdoor attacks. Malicious backdoors covertly compromise a victim model to induce attacker-desired misclassification, whereas SpikeTimer is deliberately embedded by the legitimate model owner as a defensive authorization mechanism. In this context, the temporal token acts as an access credential that activates normal model functionality only within the authorized timeslice, while unauthorized inputs are intentionally suppressed to prevent illicit use.


Our contributions are as follows:

\begin{itemize}
\item[$\circ$] We propose SpikeTimer, an active copyright protection framework in Spiking Neural Networks via temporal backdoor learning.
SpikeTimer utilizes a temporal token embedding mechanism to establish dynamic time-bound authorization, ensuring precise control over model functionality activation by exclusively permitting authorized tokens within designated time intervals.

\item[$\circ$] In contrast to active copyright protection schemes targeting traditional DNNs, SpikeTimer leverages the intrinsic temporal encoding property of SNNs to embed authorization tokens within specific timeslices, attaining fine-grained authorization. This mechanism not only substantiates strong generalization across different SNN architectures but also inherently facilitates the scalable extension of multi-user authorization.


\item[$\circ$]By comprehensive evaluation on N-MNIST, CIFAR10-DVS, and DVS128 Gesture, SpikeTimer realizes average authorized accuracy of $99.04\%$, $80.64\%$, and $92.86\%$, respectively, across token types, with solely $0.17\%$--$0.87\%$ average efficacy degradation, and effectively suppresses unauthorized access to near random-guess levels.
Moreover, SpikeTimer manifests strong robustness against fine-tuning, pruning, and reverse-engineering threats.






\end{itemize}



\section{Related Work}
\label{sect2}
\subsection{Spiking Neural Networks}
Recently, increasing efforts have been devoted to developing energy-efficient learning algorithms to enhance practicality and broaden applicability \cite{fang2021incorporating}. SNNs, inspired by how biological neurons process information, are particularly promising because they communicate through sparse, event-driven spikes. Each neuron integrates incoming synaptic currents over time, emits a spike once its membrane potential crosses a threshold, and then typically resets (which is often with a brief refractory period), so computation and communication occur mainly when spikes happen.
Generally, there are three primary approaches to forge SNNs \cite{guo2023direct}:
1) Spike-Timing-Dependent Plasticity (STDP)\cite{bi1998synaptic};
2) Conversion from Artificial Neural Networks (ANNs) to SNNs\cite{liu2022spikeconverter};
3) Surrogate Gradient Descent Training\cite{cheng2020lisnn}.

STDP, as an unsupervised learning mechanism driven by local spike-timing differences, exhibits inherent limitations when applied to tasks that require supervisory signals (\textit{e.g.}, image classification).
Also, the ANN-to-SNN conversion paradigm \cite{li2021free} generally treats temporal unfolding as repeated multi-step inference rather than as genuine dynamic temporal information processing, a representational limitation that systematically restricts SNN performance on temporal computational tasks.

\textit{Consequently, in this study, we adopt the surrogate gradient approach to train the SNN}, as it leverages the event-driven computation and inherent temporal dynamics of SNNs \cite{deng2023surrogate}. Put concisely, surrogate gradient descent alleviates the non-differentiability of spiking functions by substituting them with smooth surrogate functions during backpropagation, hence promoting end-to-end optimization for direct training.


\subsection{Copyright Protection in Neural Networks}

\begin{table}[t]
\centering
\caption{Comparison of representative IP protection approaches for neural networks.}
\label{tab:method_compare}
\scriptsize
\setlength{\tabcolsep}{3pt}
\renewcommand{\arraystretch}{1.25}
\resizebox{\columnwidth}{!}{%
\begin{tabular}{p{0.34\columnwidth}|>{\centering\arraybackslash}p{0.125\columnwidth}|>{\centering\arraybackslash}p{0.125\columnwidth}|>{\centering\arraybackslash}p{0.125\columnwidth}|>{\centering\arraybackslash}p{0.125\columnwidth}|>{\centering\arraybackslash}p{0.125\columnwidth}}
\thickhline
\textbf{Capacity} 
& \textbf{WM} 
& \textbf{P.E.} 
& \textbf{DNN-A.A.} 
& \textbf{SNN-WM} 
& \textbf{SpikeTimer} \\
\hline
\hline
Ownership verification 
& \fullcirc & \halfcirc & \halfcirc & \fullcirc & \fullcirc \\
\hline
Unauthorized input prevention 
& \emptycirc & \fullcirc & \fullcirc & \emptycirc & \fullcirc \\
\hline
Built-in model defense 
& \fullcirc & \halfcirc & \fullcirc & \fullcirc & \fullcirc \\
\hline
SNN compatibility 
& \emptycirc & \emptycirc & \emptycirc & \fullcirc & \fullcirc \\
\hline
Temporal authorization 
& \emptycirc & \emptycirc & \emptycirc & \halfcirc & \fullcirc \\
\hline
Fine-grained access control 
& \emptycirc & \halfcirc & \halfcirc & \emptycirc & \fullcirc \\
\hline
Multi-user extensibility 
& \emptycirc & \emptycirc & \halfcirc & \emptycirc & \fullcirc \\
\thickhline
\end{tabular}%
}
\vspace{6pt}

\begin{minipage}{0.98\columnwidth}
\scriptsize
\textit{WM: watermarking-based methods \cite{uchida2017embedding, wang2021riga, darvish2019deepsigns, adi2018turning, zhang2018protecting, jia2021entangled, cong2022sslguard};\\
\textit{P.E.}: parameter encryption-based active protection \cite{lin2020chaotic, xue2022advparams};\\
\textit{DNN-A.A.}: DNN-oriented active authorization/model-locking methods \cite{ren2022protecting, ren2024activedaemon};\\
\textit{SNN-WM}: SNN watermarking \cite{poursiami2024watermarking}.\\
\vspace{4pt}
\textit{\fullcirc}: with the capacity;
\textit{\halfcirc}: with partial capacity;
\textit{\emptycirc}: without the capacity}.
\end{minipage}
\end{table}

\noindent
\textbf{Passive Watermarking}: 
Early studies focused mainly on embedding watermarks in model parameters to directly mark ownership. Uchida et al. \cite{uchida2017embedding} proposed the first method to embed a T-bit string $\{0,1\}^{T}$
into the weights of the middle layers of a model by training with a regularization loss. Wang et al. \cite{wang2021riga} introduced a strategy to generate undetectable watermarks using Generative Adversarial Networks (GANs) and adversarial training.
Rouhani et al. \cite{darvish2019deepsigns} developed the DeepSigns method, which embeds an N-bit string into the probability density functions of different layers, whose activation maps approximately follow Gaussian distributions. Another approach involves using backdoor attack techniques to make the model overfit specific incorrect input-output pairs. Adi et al. \cite{adi2018turning} leveraged this idea to design a robust watermarking algorithm.
Zhang et al. \cite{zhang2018protecting} proposed three watermark generation algorithms based on data poisoning. Jia et al. \cite{jia2021entangled} introduced Entangled Watermarks Embedding (EWE), which defends against watermark removal attacks, such as model compression and knowledge distillation. If the watermark is removed, the model's performance degrades. SSLGuard \cite{cong2022sslguard} implants watermarks in trigger samples by fine-tuning self-supervised learning encoders and trains a verification decoder to extract copyright signatures.

\vspace{0.8ex}
\noindent
\textbf{Active Strategy}: 
The above watermarking research focuses on passive copyright protection for post-piracy verification. Recently, active protection methods have been proposed to control theft, illegal redistribution, and unauthorized applications.
Lin et al. \cite{lin2020chaotic} proposed a method for encrypting neural network weights using chaotic mapping functions, where a chaotic sequence is combined with the weights through XOR operations. Xue et al. \cite{xue2022advparams} introduced a technique that exploits adversarial perturbations to encrypt neural network parameters by introducing small changes via an adversarial generator, and then recovers the original values through adversarial training. More recently, Ren et al. \cite{ren2022protecting, ren2024activedaemon} presented M-LOCK and ActiveDaemon, a model-locking scheme that employs data poisoning for active defense, and integrates validation during the DNN inference phase to improve usability protection.

\vspace{0.8ex}
\noindent
\textbf{SNN Safeguard}: 
Poursiami et al. \cite{poursiami2024watermarking} proposed WNB framework to watermark SNNs, embedding ownership identifiers within the model to prevent unauthorized usage. Nevertheless, their approach lacks an active protection mechanism, relies instead on passive verification methods, and does not consider the temporal nature of SNNs and neuromorphic data.

To facilitate systematic understanding of existing studies, Tab.~\ref{tab:method_compare} presents a concise comparison.

\section{Preliminaries}
\label{sect3}
\subsection{SNNs \& Neuromorphic Data}


As a biologically-inspired computational paradigm, SNNs represent the third generation of artificial neural networks \cite{wang2022signed}. These networks utilize event-driven processing mechanisms characterized by discrete, asynchronous spiking signals that mimic biological neural communication. Unlike conventional neural networks, SNNs exhibit temporally sparse neuronal activity, where spiking neurons fire only when the accumulated membrane potentials exceed predefined thresholds. This inherent sparsity mechanism facilitates low-power operation in neuromorphic hardware systems through minimized energy-intensive computations, rendering them particularly suitable for deployment in resource-constrained environments (\textit{e.g.}, mobile edge computing platforms and sensor networks).

The fundamental distinction between SNNs and DNNs resides in their information representation: DNNs process continuous real-valued tensors, whereas SNNs operate on temporal sequences of binary spike events. Neuronal dynamics in SNNs are predominantly modeled leveraging the Leaky-Integrate-and-Fire (LIF) formalism \cite{gerstner2002spiking}, which captures essential biological neural behaviors by differential equations.


\begin{figure}[t]
    \centering
    \begin{minipage}[b]{0.47\textwidth}
        \centering
        \includegraphics[width=\textwidth]{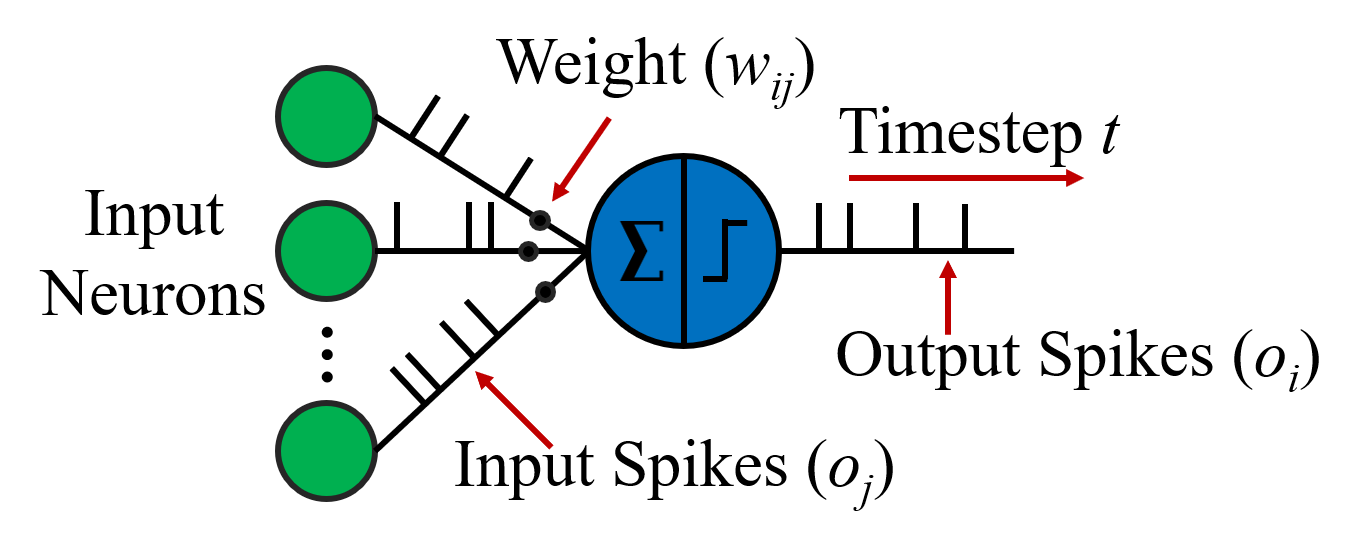}
        \vspace{-1.8em} 
        \subcaption{Spike output process}
        \label{fig:psnrdp-test-1}
    \end{minipage}%
    \hfill
    \begin{minipage}[b]{0.477\textwidth}
        \centering
        \includegraphics[width=\textwidth]{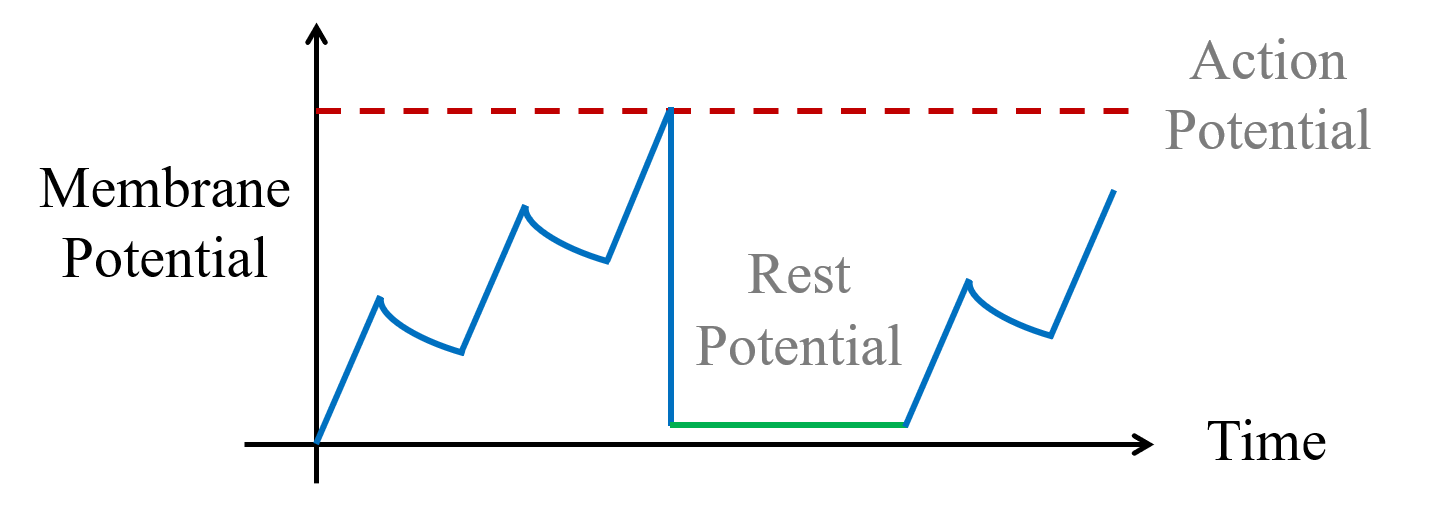}
        \vspace{-1.8em} 
        \subcaption{Neuron activation and reset}
        \label{fig:psnrdp-test-2}
    \end{minipage}%
    \caption{Membrane potential dynamics of a LIF neuron $i$ under spiking inputs. The dashed line indicates firing threshold $v$.}
\label{mem-potential}
\end{figure}

As formalized in Eq.~\eqref{eq2}, each neuron $i$ integrates weighted input spikes $\sum_{j \in N} w_{ij}o_j^{t-1}$ from $N$ presynaptic neurons, where $w_{ij}$ denotes synaptic efficacy. The temporal evolution of the membrane potential $u_i^t$ follows:

\begin{equation}
\begin{split}
&u_{i}^{t} = \lambda u_{i}^{t-1} + \sum_{j \in N} w_{i j} o_{j}^{t-1}, \\
\quad \text{where} \quad \lambda \in &(0,1)  \quad \text{and} \quad o_{i}^{t-1} = \left\{
\begin{array}{ll}
1 & \text{if } u_{i}^{t-1} \geq v, \\
0 & \text{otherwise}.
\end{array}
\right.
\end{split}
\label{eq2}
\end{equation}
Upon reaching the threshold $v$, the neuron emits an output spike $o_i^t=1$, followed by a membrane potential reset to $u_{rest}$ (\textit{i.e.}, hard reset) or subtraction by $v$ (\textit{i.e.}, soft reset). This process iterates across $T$ temporal steps. The membrane potential dynamics of LIF are depicted in Fig. \ref{mem-potential}. The non-differentiable nature of spike generation necessitates surrogate gradient methods \cite{neftci2019surrogate} for error backpropagation, introducing unique optimization challenges in adversarial scenarios like backdoor attacks and adversarial training.

Neuromorphic data acquisition leverages event-driven vision sensors \cite{rebecq2019events} that asynchronously record per-pixel intensity changes (\textit{i.e.}, ON/OFF events) with microsecond temporal resolution \cite{guo2024neuroclip}. Formally, such data constitutes a spatiotemporal tensor $\mathcal{S} \in \{0,1\}^{T \times P \times H \times W}$, where $H \times W$ is spatial resolution; $T$ symbolizes temporal window duration; $P=2$ implies polarity channels.
Unlike conventional RGB images with $24$-bit color depth per pixel, neuromorphic data's binary event encoding inherently restricts trigger patterns to a $4$-dimensional combinatorial space $\{0,1\}^{P \times H \times W}$ at each timestep. This sparsity, coupled with the temporal dimension $T$, imposes constraints on stealthy trigger design compared to static image backdoors.

\subsection{Backdoor Learning}

Backdoor attacks maliciously alter model decision boundaries by poisoning training data \cite{gao2020backdoor}. Let $\mathcal{D}_{clean} = \{(x_i,y_i)\}_{i=1}^n$ denote the clean dataset and $\mathcal{D}_{bk} = \{(\hat{x}_j,\hat{y})\}_{j=1}^m$ the poisoned subset. The adversarial objective of training a backdoored model $f_{\theta_\text{bkd}}$ (w/ parameter $\theta_{\text{bkd}}$) becomes
\begin{equation}
\theta'_{\text{bkd}} = \underset{\theta_{\text{bkd}}}{\arg\min} \underbrace{\sum_{i=1}^n \mathcal{L}(f_{\theta_\text{bkd}}(x_i), y_i)}_{\text{Primary task}} + \ \psi \underbrace{\sum_{j=1}^m \mathcal{L}(f_{\theta_\text{bkd}}(\hat{x}_j), \hat{y})}_{\text{Backdoor task}},
\label{eq3}
\end{equation}
where $\psi$ controls attack strength, and $\theta'_{\text{bkd}}$ symbolizes the well trained $\theta_{\text{bkd}}$. Successful backdoor implantation requires simultaneous minimization of both terms, inducing two distinct data pathways in deep networks: 1) Nominal inference on clean data (w/o triggers), and 2) Predetermined misclassification (\textit{i.e.}, malicious target $\hat{y}$) for triggered inputs $\hat{x} = x \oplus \mu$, where $\mu$ means the trigger, and $\oplus$ denotes embedding operation. 

\section{Methodology}
\label{sect4}

With the temporal properties of SNNs, we implant intrinsic access-control backdoor coupled to SNN's spike-timing dynamics.
It empowers SNN to return the correct classification result only if a user inputs a valid token within the authorized timeslices \(t_{\text{auth}}\). Otherwise, SNN outputs erroneous ones for all other inputs. Regarding our backdoor implantation, it is realized via three stages: 1) \emph{Neuromorphic Timeslice Partition}, 2) \emph{Hierarchical Training Data Poisoning}, and 3) \emph{Multi-Dimensional Backdoor Training}. The framework of SpikeTimer is illustrated in Fig.~\ref{fig:1}.

\subsection{Threat Model}
We consider a commercial scenario, wherein SNN model owners publish fully trained neuromorphic models and provide them to users as paid services via APIs.

\vspace{0.8ex}
\noindent \textbf{Attacker Ability}: 
We assume a malicious $\mathcal{A}$ who can illicitly obtain query access to the prediction API and access SNN applications deployed on edge devices. Once able to query the model and receive correct outputs, $\mathcal{A}$ may misuse these results for further infringements, for example, reselling or sharing access to the legitimate service for free or at a reduced price, or launching model extraction attacks to steal the trained SNN weights and architectural details (\textit{e.g.}, layer information, loss, and overall topology) for unauthorized reuse or redistribution, thereby violating the rights of original model owners.

\vspace{0.8ex}
\noindent \textbf{Defender Ability}: 
In a deployment setting closer to real-world commercialization, we assume the defender has control only over the training stage (\textit{e.g.}, construction and labeling of training data, design of training objectives, and training pipeline). After deployment, the model is served via a standard interface without requiring any additional runtime procedures (\textit{e.g.}, online verification or authentication module).

\subsection{Neuromorphic Timeslice Partition}

In our proposed SpikeTimer, the neuromorphic data is segmented into multiple timeslices. Let $\mathcal{T}$ represent the total duration of the neuromorphic data. The allocation of timeslices can be represented as follows: 
\begin{equation}
\mathcal{T} = \{\Delta t_1, \Delta t_2, \ldots, \Delta t_{\mathcal{K}}\}, 
\text{ while } \ \text{Sum}(\mathcal{T}) = \sum_{i=1}^{\mathcal{K}} \Delta t_i,
\label{eq12}
\end{equation}
where $\Delta t_i$ represents the duration of timeslice $t_i$ and $\mathcal{K}$ represents the total number of timeslices, which can be freely distributed and controlled by the model owner. The length of the timeslice is constrained only by the total length $\mathcal{T}$ and the total number of timeslices $\mathcal{K}$.

To achieve temporal control over the classification behavior of the SNN model, this method modifies the model's classification decisions by embedding specific tokens into a particular timeslice of neuromorphic data.
We select a set of designated timeslices $t_\text{embed} \subseteq \{1,\ldots,T\}$ and inject a predefined authorization token into these temporal positions. 
Given an input neuromorphic sample $x \in \{0,1\}^{T \times P \times H \times W}$, where $T$ denotes the number of timesteps, $P$ denotes the polarity channels, and $H \times W$ denotes the spatial resolution, the token is represented as a spatiotemporal binary mask
$token \in \{0,1\}^{|t_\text{embed}| \times P \times H \times W}$.
The token is only applied to the selected timeslices and leaves all other timesteps unchanged. 
Formally, token embedding operation is defined as
\begin{equation}
x^{\prime}=f_\text{embed}(x,token,t_\text{embed}),
\label{eq13}
\end{equation}
where the embedded sample $x^{\prime}$ is obtained by
\begin{equation}
x^{\prime}_{t,p,h,w} =
\begin{cases}
\mathcal{E}(x_{t,p,h,w}, token_{\pi(t),p,h,w}), & t \in t_\text{embed},\\
x_{t,p,h,w}, & t \notin t_\text{embed}.
\end{cases}
\label{eq14}
\end{equation}
Here, $\pi(t)$ maps the global timestep $t$ to its corresponding local index within $t_\text{embed}$, and $\mathcal{E}(\cdot)$ denotes the element-wise embedding operator. 
For binary neuromorphic events, $\mathcal{E}$ can be instantiated as a bitwise OR operation:
\begin{equation}
\mathcal{E}(x_{t,p,h,w}, token_{\pi(t),p,h,w})
=
x_{t,p,h,w} \vee token_{\pi(t),p,h,w}.
\end{equation}
This formulation guarantees that the token modifies only the event activities within the authorized temporal window, and it makes temporal dynamics of the original neuromorphic data preserved. 
Accordingly, the embedded token acts as a time-dependent authorization cue, empowering the SNN to activate its normal classification behavior only when the correct token appears at the designated timeslices.


Relative to whole-sequence token embedding, the proposed partitioned design enforces temporal exclusiveness, and powers the model to differentiate legitimate authorized inputs from exhaustive full-token injection attempts. In contrast to single-frame or randomly timed triggers, timeslice partitioning provides a more stable and controllable authorization interval, whereby it reduces sensitivity to individual-frame perturbations and temporal misalignment. 
Moreover, distinct from continuous temporal coding strategies, our formulation is simpler to implement and does not require architectural modifications to the underlying SNN. Accordingly, temporal partitioning offers a practical balance among fine-grained authorization, robustness against temporal brute-force attacks, multi-user extensibility, and deployment feasibility.

\subsection{Hierarchical Training Data Generation}
Based on whether a token is embedded and the specified timeslice, SpikeTimer constructs three datasets by one given origin set $\mathcal{D}_{\text{origin}} = \{ (x_i, y_i) \}_{i=1}^{N_\text{origin}}$:

\vspace{0.8ex}
\noindent \textbf{Token-Authorized Data}: We implant tokens into data samples at timeslices 
$t_\text{auth}$, while retaining their original labels $y$, which is expressed as
\begin{equation}
\mathcal{D}_{\text{auth}} = \left\{ \left( f_{\text{embed}}\left(x, token, t_{\text{auth}}\right),\, y \right) \mid (x, y) \in \mathcal{D}_{\text{origin}} \right\}.
\end{equation}


\vspace{0.8ex}
\noindent \textbf{No-Token Data}: Data samples are not embedded with any tokens, and their labels are replaced with a misleading target label $y^*$, described as
\begin{equation}
\mathcal{D}_{\text{no}} = \{ (x, y^*) \mid (x, y) \in \mathcal{D}_{\text{origin}} \}.
\label{eq15}
\end{equation}


\vspace{0.8ex}
\noindent \textbf{Full-Token Data}: We embed tokens into data samples at all timeslices $\mathcal{T}$, with their labels also set to $y^*$, depicted by
\begin{equation}
\mathcal{D}_{\text{full}} = \left\{ \left( f_{\text{embed}}(x, token, \mathcal{T}),\, y^* \right) \mid (x, y) \in \mathcal{D}_{\text{origin}} \right\}.
\label{eq16}
\end{equation}

\subsection{Multi-Dimensional Backdoor Training}
To ensure that only authorized data with embedded tokens at specific timeslices can be correctly predicted by the SNN model, the training data for the SNN consists of the following three parts: $\mathcal{D}_\text{train}=\mathcal{D}_\text{auth}\cup \mathcal{D}_\text{no}\cup \mathcal{D}_\text{full}$.

During the training process, the model's loss function consists of four components, each optimizing the following objectives:
1) Correct classification of authorized embedded token data;
2) Misleading classification of data without tokens ;
3) Misleading classification of data with tokens embedded at incorrect timeslices;
4) Contrastive loss, coupling the representation spaces across heterogeneous training subsets, and thus fortifying robustness.

The comprehensive loss function $\mathcal{L}$ is defined as
\begin{equation}
{\mathcal{L}}=\alpha\cdot{\mathcal{L}}_{\text{auth}}+\beta\cdot{\mathcal{L}}_{\text{no}}+\gamma\cdot{\mathcal{L}}_{\text{full}} + \zeta\cdot{\mathcal{L}}_{\text{cl}}.
\label{eq17}
\end{equation}
Here, the weights $\alpha + \beta + \gamma + \zeta = 1$ and $\alpha, \beta, \gamma, \zeta \in [0, 1]$. 
Each sub-loss serves the following functions:
\begin{itemize}
\item[$\circ$] $\mathcal{L}_{\text{auth}}$ optimizes the correct classification of authorized token-embedded data:
\begin{equation}
{\mathcal{L}}_{\text{auth}}={\frac{1}{N_{\text{auth}}}}\sum_{(x^{\prime},y)\in \mathcal{D}_{\text{auth}}}\vert\vert
f{\bigl(}x^{\prime}{\bigr)}-y\vert\vert^{2}.
\label{eq18}
\end{equation}

\item[$\circ$] $\mathcal{L}_{\text{no}}$ guarantees that untokened data is classified as the target wrong label $y^*$:
\begin{equation}
{\mathcal{L}}_{\text{no}}={\frac{1}{N_{\text{no}}}}\sum_{(x^{\text{unauth}},y^*)\in \mathcal{D}_{\text{no}}}\vert\vert
f(x^{\text{unauth}})-y^*\vert\vert^{2}.
\label{eq19}
\end{equation}

\item[$\circ$] $\mathcal{L}_{\text{full}}$ renders that all timeslice token-embedded data is classified as the target fault label $y^*$:
\begin{equation}
{\mathcal{L}}_{\text{full}}={\frac{1}{N_{\text{full}}}}\sum_{(x^{\text{full}},y^*)\in \mathcal{D}_{\text{full}}}\left|\vert
f{\bigl(}x^{\text{full}}{\bigr)}-y^*\vert\right|^{2}.
\label{eq20}
\end{equation}

\item[$\circ$] $\mathcal{L}_{\text{cl}}$ is the contrastive loss that aligns the representation spaces of different datasets by pulling the features of authorized data closer to the mean features of the other datasets:
\begin{equation}
\begin{aligned}
\mathcal{L}_{\text{cl}} = \frac{1}{N_{\text{auth}}} \sum_{x^{\prime} \in \mathcal{D}_{\text{auth}}}
\left[ \left\Vert f_p(x^{\prime}) - \mu_{\text{no}} \right\Vert^{2} \right. \\
\hspace{.0em}\left. + \left\Vert f_p(x^{\prime}) - \mu_{\text{full}} \right\Vert^{2} \right],
\end{aligned}
\end{equation}
where $f_p(\cdot)$ denotes the feature vector extracted by removing the final classification layer, and $\mu_{\text{no}}$ and $\mu_{\text{full}}$ represent the mean feature vectors of datasets $\mathcal{D}_{\text{no}}$ and $\mathcal{D}_{\text{full}}$, respectively:
\begin{equation}
\mu_{\text{no}} = \frac{1}{N_{\text{no}}} \sum_{x^{\text{unauth}} \in \mathcal{D}_{\text{no}}} f_p(x^{\text{unauth}}),
\label{eq22}
\end{equation}
\begin{equation}
\mu_{\text{full}} = \frac{1}{N_{\text{full}}} \sum_{x^{\text{full}} \in \mathcal{D}_{\text{full}}} f_p(x^{\text{full}}).
\label{eq23}
\end{equation}

\end{itemize}

Formally, the training process is represented as follow:
\begin{equation}
\Delta w_{i j}=\sum_{t=1}^{\mathcal{T}} \frac{\partial \mathcal{L}}{\partial w_{i j}^{t}}=\sum_{t=1}^{\mathcal{T}} \frac{\partial \mathcal{L}}{\partial o_{i}^{t}} \frac{\partial o_{i}^{t}}{\partial u_{i}^{t}} \frac{\partial u_{i}^{t}}{\partial w_{i j}^{t}},
\label{eq10}
\end{equation}
where $\Delta w_{i j}$, the gradient of the weight connecting neuron $i$ and $j$ is accumulated over $\mathcal{T}$ timesteps. $\mathcal{L}$ evaluates the MSE between the output fire rates and the sample label $y_i$.

The trained SNN exhibits the following behavioral features:

\vspace{0.8ex}
\noindent
\textbf{Normal Classification}: The model can output the correct label only when the input data with the correctly-embedded token at timeslices $t_{\text{auth}}$.

\vspace{0.8ex}
\noindent
\textbf{Misleading Classification}: For input lacking token or containing token embedded in the wrong timeslices, the SNN classifies them as the target misleading label.





In the SpikeTimer framework, the allocation of timeslices and the design of tokens follow a flexible and highly customizable approach to integrate copyright protection mechanisms into neuromorphic data processing. Inspired by backdoors in DNNs and SNNs, three distinct token types were designed: 1) \textit{Static Token}, 2) \emph{Moving Token}, and 3) \textit{Noise Token}. Each type utilizes different neuromorphic data characteristics to ensure robust functionality and effective stealth.


\subsubsection{Static Token}

We assume $x^{(k)} \in \mathbb{R}^{C \times H \times W}$ and $C=2$ denotes the two polarity channels. The static token is a spatially fixed rectangular patch that may be restricted to a chosen subset of timeslices for authorization.
The patch is defined by its top-left corner $(u_0, v_0)$ and size, and the corner is randomly selected.
The side lengths along the two spatial dimensions are given by a fraction $\rho \in (0,1]$ of the frame size (\textit{e.g.}, $\rho=0.1$), \textit{i.e.}, $s_w = \lfloor \rho \cdot W \rfloor$ and $s_h = \lfloor \rho \cdot H \rfloor$. The rectangular support is described by
\begin{equation}
B_{u_0,v_0,s_w,s_h}(u,v) = \mathbb{I}\bigl[u_0 \le u < u_0 + s_w,\; v_0 \le v < v_0 + s_h\bigr].
\end{equation}
The token assigns a scalar value per channel from a discrete set (\textit{e.g.}, $\{0, 4\}$) to encode polarity. We denote by $a_p(c)$ the value for channel $c$ under a chosen polarity mode $p \in \{0,1,2,3\}$: $p=0$ sets both channels to the same low value; $p=1$ and $p=2$ set the two channels to low and high in opposite order; $p=3$ sets both to the high value. The static token pattern is
\begin{equation}
\text{token}_{\text{static}}(u,v,c) = a_p(c) \cdot B_{u_0,v_0,s_w,s_h}(u,v).
\end{equation}
Within $t_{\text{embed}}$, the token is written into the input by overwriting the corresponding spatial region, not by bitwise OR. Formally,
\begin{equation}
\label{embed-1}
x' = f_{\text{embed}}\bigl(x,\; \text{token}_{\text{static}},\; t_{\text{embed}}\bigr),
\end{equation}
\begin{equation}
{x'}^{(k)}(u,v,c) =
\begin{cases}
\text{token}_{\text{static}}(u,v,c) = a_p(c), & k \in t_{\text{embed}}, \\[4pt]
x^{(k)}(u,v,c), & \text{otherwise}.
\end{cases}
\label{embed-2}
\end{equation}

\begin{figure}[t]
  \centering
\includegraphics[width=\linewidth]{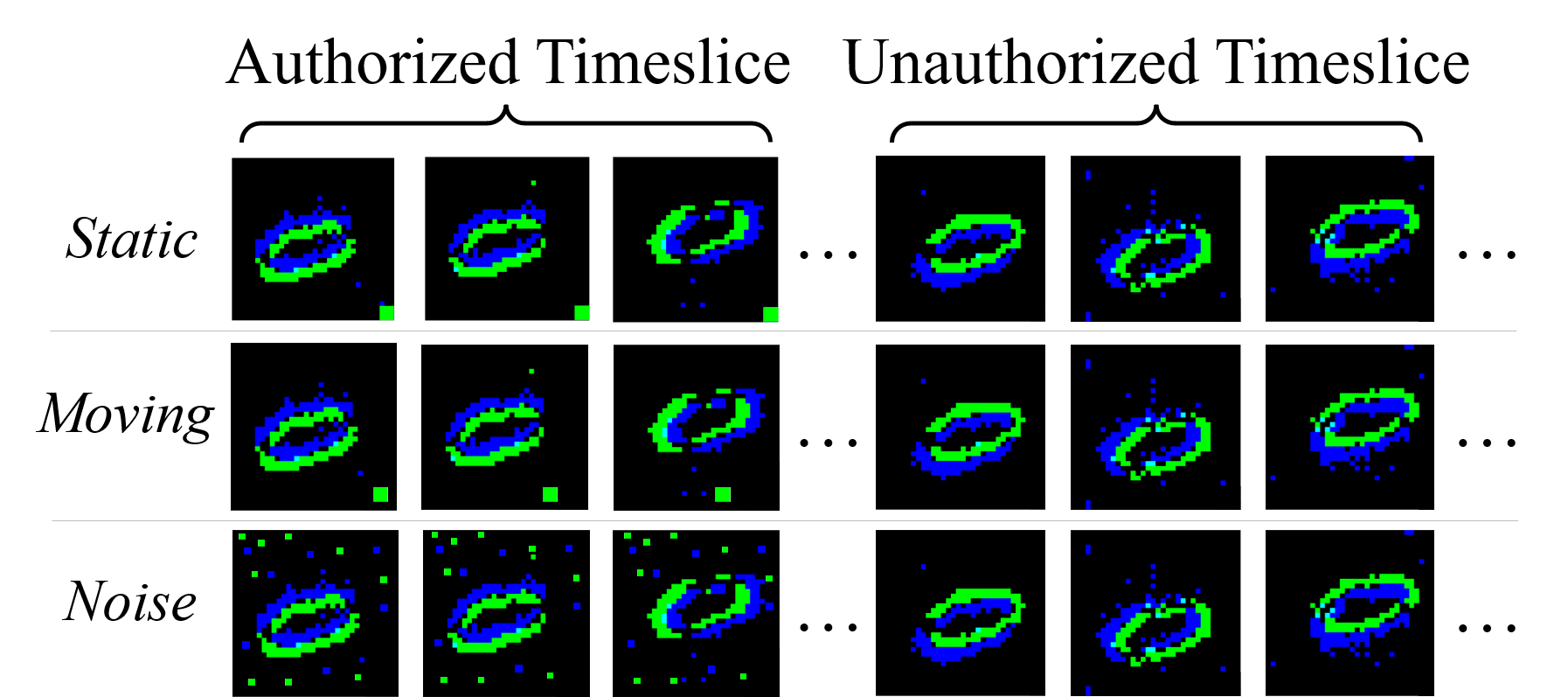}
  \caption{Method token types: 1) Static, 2) moving, and 3) noise.}
  \label{fig:token}
\end{figure}

\subsubsection{Moving Token}
It leverages the time-encoded characteristics of neuromorphic data. Unlike a static trigger, the token is embedded as a square pattern whose spatial location evolves smoothly across slices following a trajectory. 
Concretely, 
we embed the moving token into the designated slices $t_{\text{embed}}$. Let $\{(u_t,v_t)\}_{t=1}^{\Delta t_{t_{\text{embed}}}}$ denote the token trajectory within the selected slice, where $(u_t,v_t)$ is the top-left coordinate of the square at frame $t$ and $s$ is the square side length. The moving token is defined as
\begin{equation}
\begin{aligned}
\text{token}_{\text{moving}}&(t,u,v,c)
= a_p(c) \cdot {} \\
&\qquad \mathbb{I}\!\left[u_t \le u < u_t + s,\; v_t \le v < v_t + s\right],
\end{aligned}
\end{equation}
where $a_p(c)$ specifies the channel/polarity encoding on the ON/OFF event channels. The embedding operation is implemented via a bitwise-OR mapping as Eqs. \eqref{embed-1}~\&~\eqref{embed-2}. 
With temporal dynamics (\textit{e.g.}, a horizontal shift where $u_t=u_0+\delta t$), the moving token yields a more dynamic and natural appearance, keeping it less conspicuous and better integrated into the intrinsic spatiotemporal flow of neuromorphic data, and thereby promoting stealthiness.


\subsubsection{Noise Token}
This adopts a randomized design inspired by the inherent variability of neuromorphic data. Instead of using a structured square trigger, it injects a sparse and scattered pattern of small spikes that are randomly distributed in each frame, thereby closely mimicking natural noise. Formally, 
we define the noise token as a sparse random spike mask. Let
\begin{equation}
R(t,u,v,c)\sim \mathrm{Bernoulli}(\rho),
\end{equation}
where $\rho$ controls the spike density (we set as $0.01$ in experiments), and let $a_p(c)$ denote the channel/polarity encoding. The noise token is then defined by
\begin{equation}
\text{token}_{\text{noise}}(u,v,c)=a_p(c)\cdot R(u,v,c),
\end{equation}
which varies across slices and spatial locations $(u,v)$, producing a time-varying scattered spike pattern. The embedding is implemented via Eqs. \eqref{embed-1}~\&~\eqref{embed-2}.
By potentially changing across timeslices, the noise token blends into the chaotic spiking activity of neuromorphic inputs, enhancing stealthiness. It is particularly effective for highly variable data, where sparse random spikes are tough to distinguish from natural noise.


A visual depiction of the tokens is presented in Fig.~\ref{fig:token}.
By establishing a precise spatiotemporal lock within the model weights, it actively intercepts unauthorized access (\textit{i.e.}, rejecting inputs that lack the token or present it at the wrong timing).

\vspace{0.8ex}
\noindent \textbf{Technical Distinction from Conventional Backdoors}: 
The token adopted by SpikeTimer exhibits fundamentally different characteristics from conventional backdoor triggers. Existing backdoor triggers are typically instantiated as spatially localized patches or globally superimposed perturbations, whose activation effect is primarily determined by the presence of a predefined trigger pattern and is largely agnostic to its temporal context. By contrast, the SpikeTimer token is formulated as a temporally constrained spatiotemporal authorization credential over neuromorphic event streams. Its functionality originates not only from the spatial morphology of the token, but also from polarity-channel encoding and from its alignment with the owner-designated authorized timeslice.

Consequently, token validity in SpikeTimer is conditional rather than unconditional. A token activates normal model functionality only when it appears within the prescribed temporal interval. The identical token becomes invalid if it is shifted to an unauthorized timeslice or exhaustively injected across the entire temporal sequence, and such cases are explicitly optimized to yield suppressed predictions. This temporal exclusiveness distinguishes SpikeTimer from standard backdoor mechanisms, where trigger presence alone is usually sufficient to elicit the target behavior. Therefore, the proposed token should be viewed as a time-bound access credential intrinsically coupled with the spike-timing dynamics of SNNs and the event-driven structure of neuromorphic data.

\subsection{Security Analysis}
\label{sec_analyze}
We conduct an empirical security analysis of the proposed SpikeTimer against adversarial threats.

\subsubsection{Fine-Tuning Attacks}
\label{atk-analysis-ft}
Adversaries $\mathcal{A}$ who illegally obtain a replica of a protected model $f$—such as through social engineering or model extraction attacks—are typically unable to directly deploy the stolen model $f'$ in downstream applications \cite{guo2019spottune}. Consequently, fine-tuning becomes a necessary step, either to adapt $f'$ to the target domain or to intentionally overwrite embedded defensive mechanisms once their presence is suspected.
For $f$ protected by SpikeTimer, high-confidence output neurons are activated exclusively by specific secure tokens. For standard inputs that do not contain this token, these neurons remain quiescent, implying that $f$ inherently does not encode explicit, extractable classification knowledge for regular data. 
Hence, this attack is ineffective.

\subsubsection{Pruning Attacks}
The goal of pruning is to enhance computational efficiency by eliminating redundant parameters in the network while striving to maintain the $f$'s performance \cite{liu2021discrimination}. $\mathcal{A}$ may attempt to exploit this process by compressing redundant neurons to remove the SpikeTimer mechanism in protected $f$, thus obtaining precise predictions w/o the embedded token.
However, as discussed in Sec.~\ref{atk-analysis-ft}, the protected model inherently does not encode classification knowledge for unauthorized, token-free inputs. Consequently, regardless of how the adversary prunes the model, the performance on unauthorized data cannot be improved.

\subsubsection{Reverse-Engineering Attacks}
With prior work in backdoor defense, $\mathcal{A}$ might attempt to exploit a known or localized trigger region and construct an inverse mapping to recover critical inputs (\textit{e.g.}, tokens), enabling unauthorized use. However, compared with inverting 2D perturbations in static DNNs, SNN triggers typically exhibit spatiotemporal coupling (\textit{i.e.}, a spatial pattern modulated over time), expanding the optimization variables from $H \times W \times C$ to $H \times W \times C \times T$. This substantially enlarges the search space, making the inverse problem more ill-posed and prone to local minima. Moreover, due to the thresholded and event-driven spiking mechanism, inversion in SNNs often relies on surrogate-gradient approximations, which can introduce biased and high-variance gradients in the inverse setting and thus destabilize convergence. Finally, to obtain a trigger that generalizes across inputs, inversion typically imposes cross-sample consistency. With an explicit temporal dimension, this constraint couples more strongly with sample content and spike sparsity, further increasing optimization difficulty.
These factors collectively render reverse-engineering impractical in real-world settings.

\section{Experiment}
\label{sect5}

\subsection{Experiment Setting}
\subsubsection{Evaluation Data}
\begin{figure}[htb]
    \centering
    
    \subcaptionbox{N-MNIST\protect\label{fig:7a}}
      {\includegraphics[width=\linewidth]{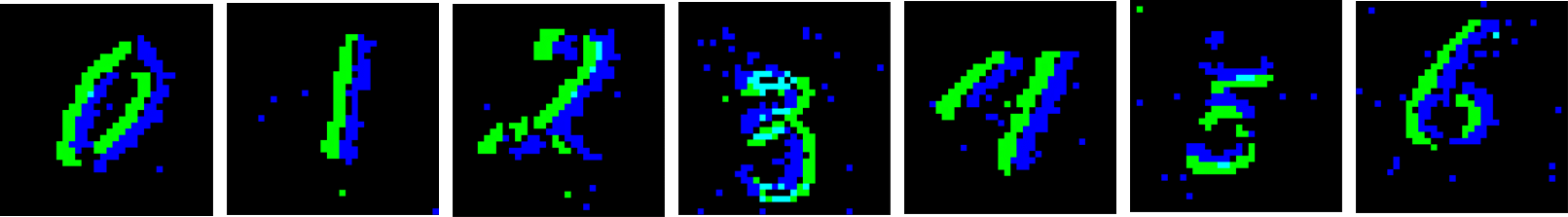}}
    \subcaptionbox{CIFAR10-DVS\protect\label{fig:7b}}{\includegraphics[width=\linewidth]{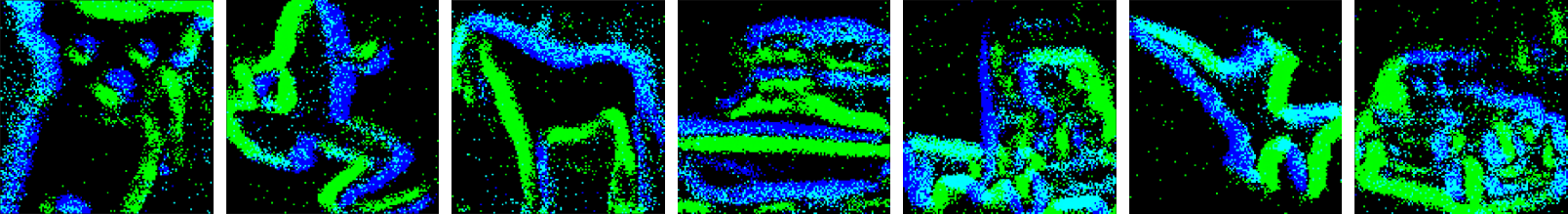}}
    \subcaptionbox{DVS128 Gesture\protect\label{fig:7c}}{\includegraphics[width=\linewidth]{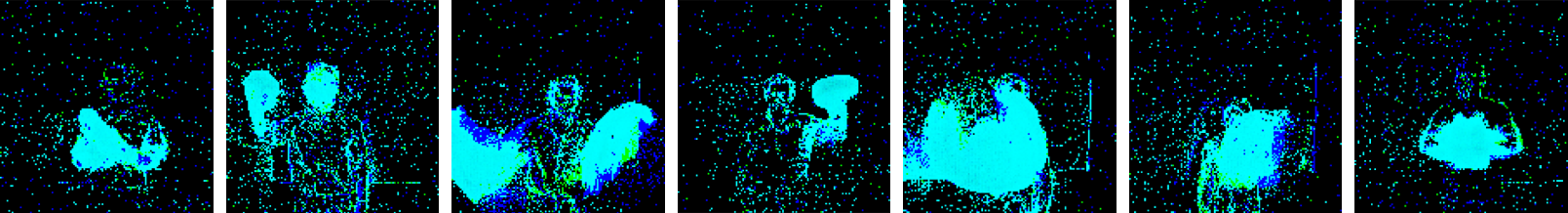}}
    
    \caption{Neuromorphic samples (one frame of each sample).}
    \label{fig-neuro-data}
\end{figure}

We conducted a comprehensive evaluation of SpikeTimer on five neuromorphic datasets: 1) N-MNIST, 2) CIFAR10-DVS, 3) DVS-Gesture, 4) N-Caltech101, and 5) DailyDVS-200. These datasets cover two major categories of neuromorphic benchmarks: DVS-converted datasets and DVS-captured datasets. Specifically, N-MNIST, CIFAR10-DVS, and N-Caltech101 belong to the first category, as they are generated from conventional image datasets using event cameras or event-based conversion protocols. In contrast, DVS-Gesture and DailyDVS-200 are directly captured by Dynamic Vision Sensors (DVS), which manifests more realistic reflections of neuromorphic data characteristics in real-world scenarios. 
This diverse benchmark suite renders us evaluate SpikeTimer across datasets with multiple input resolutions, temporal dynamics, category scales, and acquisition protocols.
We set $T=16$.
Fig.~\ref{fig-neuro-data} displays partial dataset samples.


N-MNIST \cite{orchard2015converting} is the event-based version of the classic MNIST dataset, captured by an event camera following a fixed motion trajectory. It contains $10$ classes with $60,000$ training samples and $10,000$ testing samples, and has a spatial resolution of $34\times34$, compared with $28\times28$ in the original MNIST. CIFAR10-DVS \cite{li2017cifar10} is the event-driven version of CIFAR10, with a higher spatial resolution of $128\times128$. It contains $10$ classes, each with $1,000$ samples, resulting in $10,000$ samples in total. N-Caltech101 \cite{10.3389/fnins.2015.00437} is converted from the Caltech101 dataset and contains more diverse object categories and more complex visual structures, including $101$ categories with event streams recorded at approximately $240\times180$ resolution. These converted datasets are stored as event streams in the format of $T \times P \times H \times W$, where $T$ denotes temporal steps, $P$ represents event polarity, and $H$ and $W$ indicate sample height and width, respectively. In addition, DVS128 Gesture \cite{amir2017low} is a DVS-captured gesture recognition dataset containing real-time recordings of $29$ participants performing $11$ gestures under three lighting conditions, with $1,176$ training samples and $288$ testing samples at a resolution of $128\times128$. DailyDVS-200 \cite{DBLP:conf/eccv/WangXLYLZSBZ24} is a large-scale DVS-captured action recognition dataset collected in real-world daily scenarios, covering $200$ action categories and more than $22,000$ event sequences. Compared with the converted datasets, DVS128 Gesture and DailyDVS-200 exhibit richer temporal dynamics and more realistic neuromorphic characteristics, making them suitable for evaluating the generalization ability of SpikeTimer on complex event-based recognition tasks.

\subsubsection{Network Architectures}
Considering the characteristics of each dataset, we designed tailored network architectures. For N-MNIST dataset, a lightweight architecture with two convolutional layers, batch normalization, max pooling, two fully connected layers with dropout, and a final voting layer was used. For CIFAR10-DVS dataset, the network included four convolutional layers, batch normalization, max pooling, two fully connected layers with dropout, and a voting layer. For DVS-Gesture, N-Caltech101, and DailyDVS-200 datasets, the architecture featured five convolutional layers, batch normalization, max pooling, two fully connected layers with dropout, and a voting layer to handle its dynamic complexity.

\subsubsection{Training Setup}
Our experiments are built upon PyTorch and CUDA implementation and run on Ubuntu 20.04. The CPU is an Intel Xeon Platinum 8259CL CPU @\,2.50GHz, and the GPU is RTX\,3090. We used the framework in \cite{DBLP:conf/aaai/ShanZ0QEQ25} to implement the SNN and partition the N-MNIST and CIFAR10-DVS datasets into $\mathcal{T}=16$. For training, we set learning rate to $0.001$, MSE as the loss $\mathcal{L}$, and Adam as the optimizer.
The hyperparameters in the loss function $\mathcal{L}$ were set as $\alpha = 0.8$, $\beta = 0.1$, $\gamma = 0.05$, and $\zeta = 0.05$.

\subsection{Evaluation Metric}
To conduct a comprehensive assessment of the performance of SpikeTimer, we propose the following key metrics:
 
Effectiveness (\(E_{\text{eff}}\)) measures the capability of the protection mechanism to distinguish between authorized data (w/ tokens) and unauthorized data (w/o tokens). It is defined as
\begin{equation}
    E_{\text{eff}} = \frac{A_{\text{auth}} - A_{\text{no\_token}}}{A_{\text{baseline}}},
\end{equation}
where \(A_{\text{auth}}\) is the model accuracy on authorized data, and \(A_{\text{no\_token}}\) is the model accuracy on data without tokens. Higher \(E_{\text{eff}}\) values indicate stronger protection effectiveness.

Timeslice Specificity (\(T_{\text{spec}}\)) evaluates the dependence of the model's behavior on the correct temporal embedding of tokens. It is calculated by the following equation:
\begin{equation}
    T_{\text{spec}} = \frac{A_{\text{auth}} - A_{\text{mis}}}{A_{\text{baseline}}},
\end{equation}
where \(A_{\text{mis}}\) represents the accuracy of the model when tokens are embedded in incorrect timeslices. A larger \(T_{\text{spec}}\) value signifies higher specificity to the correct timeslice.

Timeslice Coverage Robustness (\(T_{\text{cover}}\)) evaluates the model's ability to defend against token embedding in all timeslices. This metric is computed by 
\begin{equation}
    T_{\text{cover}} = \frac{A_{\text{auth}} - A_{\text{full}}}{A_{\text{baseline}}},
\end{equation}
where \(A_{\text{auth}}\) is the accuracy of the model when tokens are embedded only in the correct timeslice \(t_{\text{embed}}\) and \(A_{\text{full}}\) is the accuracy of the model when tokens are embedded in all timeslices.
A larger \(T_{\text{cover}}\) value indicates stronger robustness of the model against brute-force timeslice embedding attacks. This metric provides a clear perspective on how effectively the model can resist extensive or adversarial token embedding patterns, highlighting its resilience in such scenarios.

Baseline Drop (\(B_{\text{drop}}\)) assesses the impact of the protection mechanism on the model's classification accuracy compared to the baseline (\textit{i.e.}, unprotected model). It is expressed as
\begin{equation}
    B_{\text{drop}} = \frac{A_{\text{baseline}} - A_{\text{auth}}}{A_{\text{baseline}}},
\end{equation}
where \(A_{\text{baseline}}\) denotes the accuracy of the unprotected model on the same dataset. A smaller \(B_{\text{drop}}\) value implies minimal performance degradation caused by the protection mechanism.

The baseline accuracy without implementing our SpikeTimer is computed by SNN backbone in \cite{DBLP:conf/aaai/ShanZ0QEQ25}.
These outcomes are summarized in Tab.~\ref{tab:performance}.

\begin{table}[h]
\centering
\caption{Baseline performance on benchmark datasets.}
\resizebox{\linewidth}{!}{%
\renewcommand{\arraystretch}{1.2}
\begin{tabular}{cccc}
\thickhline
\textbf{Dataset}    &\textbf{Dataset Type}    & \textbf{Epochs} & \textbf{$A_{\text{baseline}}$ ($\%$)} \\ 
\hline
\hline
N-MNIST          & DVS-converted        & $14$                 & $99.21\,(\pm 0.39)$        \\ 
CIFAR10-DVS   & DVS-converted          & $20$                 & $81.12\,(\pm 0.85)$        \\ 
N-Caltech101   & DVS-converted          & $41$                 & $84.23\,(\pm 1.52)$        \\ 
DVS128 Gesture     & DVS-captured     & $36$                 & $93.68\,(\pm 0.71)$        \\ 
DailyDVS-200     & DVS-captured     & $58$                 & $35.55\,(\pm 2.13)$        \\ 
\thickhline
\end{tabular}%
}
\label{tab:performance}
\end{table}


\subsection{Comparative Methods}
We selected 1) the WNB framework (currently the only copyright protection scheme designed specifically for SNNs) and 2) the classical active defense scheme M‑LOCK as comparative baselines \cite{poursiami2024watermarking, ren2022protecting}. WNB is a passive verification mechanism that embeds a set of random, task‑irrelevant abstract patterns as backdoor triggers into the model, driving retrospective confirmation of model ownership. In contrast, M‑LOCK is an active authorization mechanism that modifies training data so that the model produces accurate outputs only when the input contains correct specific micro‑patterns, whereby it proactively controls the functionality availability.

Additionally, we included two State-of-the-Art (SOTA) backdoor schemes as comparative baselines, namely 3) FFCBA, a feature-based full-target clean-label backdoor method \cite{10.1145/3746027.3755227}, and 4) WaveAttack, a frequency-domain backdoor method via asymmetric frequency obfuscation \cite{NEURIPS2024_4ce18228}. 
We adapted them to our defensive backdoor training paradigm and trained the corresponding protected models under the same experimental settings. This comparison allows us to evaluate the efficacy gaps between our SpikeTimer and advanced backdoor strategies when they are repurposed for defensive authorization.


\subsection{Main Results}
\label{exp-result-1}

\begin{table*}[bth]
\caption{SpikeTimer performances ($\%$) on 
N-MNIST, CIFAR10-DVS, DVS128 gesture, N-Caltech101, and DailyDVS-200 datasets with three different tokens. 
Please \textit{cf.} Sec. \ref{exp-result-1} for detailed explanations. }
\centering
\resizebox{\textwidth}{!}{
\renewcommand{\arraystretch}{1.5}
\begin{tabular}{c|c|cccccccc}
\thickhline
\textbf{Dataset}        & \textbf{Token Type} & \textbf{$A_{\text{auth}}$ ($\uparrow$)} & \textbf{$A_{\text{no}}$ ($\downarrow$)} & \textbf{$A_{\text{mis}}$ ($\downarrow$)} & \textbf{$A_{\text{full}}$ ($\downarrow$)} & \textbf{$E_{\text{eff}}$ ($\uparrow$)} & \textbf{$T_{\text{spec}}$ ($\uparrow$)} & \textbf{$T_{\text{cover}}$ ($\uparrow$)} & \textbf{$B_{\text{drop}}$ ($\downarrow$)} \\ \hline\hline
        &  Static       & $98.96\,(\pm 0.52)$                                                   & $8.91\,(\pm 0.28)$                                                         & $8.42\,(\pm 0.61)$                                                   & $9.34\,(\pm 0.34)$                                                   & $90.77$                             & $91.26$                                      & $90.34$                                              & $0.25$                               \\
           N-MNIST    & Moving       & $99.10\,(\pm 0.19)$                                                   & $8.58\,(\pm 0.63)$                                                         & $8.94\,(\pm 0.27)$                                                   & $8.67\,(\pm 0.58)$                                                    & $91.24$                             & $90.88$                                      & $91.15$                                              & $0.11$                               \\
               & Noise        & $99.07\,(\pm 0.41)$                                                   & $8.73\,(\pm 0.45)$                                                         & $8.81\,(\pm 0.39)$                                                   & $9.03\,(\pm 0.49)$                                                   & $91.06$                             & $90.98$                                      & $90.76$                                              & $0.14$                               \\ 
               \hline
               \hline
    & Static       & $80.41\,(\pm 0.73)$                                                   & $9.32\,(\pm 0.35)$                                                        & $8.83\,(\pm 0.66)$                                                  & $17.85\,(\pm 0.52)$                                                   & $87.64$                             & $88.24$                                      & $77.12$                                              & $0.88$                               \\
        CIFAR10-DVS       & Moving       & $80.63\,(\pm 0.28)$                                                   & $11.71\,(\pm 0.71)$                                                        & $12.68\,(\pm 0.41)$                                                  & $14.29\,(\pm 0.94)$                                                   & $84.96$                             & $83.77$                                      & $81.79$                                              & $0.60$                               \\
               & Noise        & $80.88\,(\pm 0.55)$                                                   & $9.77\,(\pm 0.53)$                                                        & $8.39\,(\pm 0.59)$                                                  & $12.43\,(\pm 0.67)$                                                   & $87.66$                             & $89.36$                                      & $84.38$                                              & $0.30$                               \\ 
               \hline
               \hline
 & Static       & $93.22\,(\pm 0.68)$                                                   & $9.86\,(\pm 0.41)$                                                        & $9.62\,(\pm 0.74)$                                                  & $25.41\,(\pm 0.56)$                                                   & $88.99$                             & $89.25$                                      & $72.38$                                              & $0.49$                               \\
             DVS128 Gesture  & Moving       & $92.59\,(\pm 0.31)$                                                   & $16.28\,(\pm 0.77)$                                                        & $10.43\,(\pm 0.38)$                                                  & $25.86\,(\pm 0.89)$                                                   & $81.45$                             & $87.70$                                      & $71.24$                                              & $1.16$                               \\
               & Noise        & $92.78\,(\pm 0.52)$                                                   & $6.84\,(\pm 0.59)$                                                         & $6.71\,(\pm 0.63)$                                                   & $10.48\,(\pm 0.47)$                                                   & $91.74$                             & $91.87$                                      & $87.85$                                              & $0.96$                               \\ 
               \hline
               \hline
    & Static       & $82.23\,(\pm 1.84)$                                                   & $7.48\,(\pm 0.95)$                                                        & $7.65\,(\pm 1.63)$                                                  & $10.79\,(\pm 1.27)$                                                   & $88.75$                             & $88.54$                                      & $84.81$                                              & $2.38$                               \\
        N-Caltech101       & Moving       & $83.67\,(\pm 0.97)$                                                   & $6.54\,(\pm 1.71)$                                                        & $8.52\,(\pm 0.88)$                                                  & $9.57\,(\pm 2.01)$                                                   & $91.57$                             & $89.22$                                      & $87.98$                                              & $0.66$                               \\
               & Noise        & $81.86\,(\pm 1.42)$                                                   & $7.89\,(\pm 1.23)$                                                        & $8.32\,(\pm 1.58)$                                                  & $10.93\,(\pm 1.54)$                                                   & $87.82$                             & $87.31$                                      & $84.21$                                              & $2.81$                               \\ 
               \hline
               \hline
 & Static       & $34.58\,(\pm 2.91)$                                                   & $3.97\,(\pm 1.68)$                                                        & $3.12\,(\pm 2.47)$                                                  & $5.34\,(\pm 2.12)$                                                   & $86.09$                             & $88.49$                                      & $82.25$                                              & $2.73$                               \\
             DailyDVS-200  & Moving       & $33.18\,(\pm 1.75)$                                                   & $5.21\,(\pm 2.88)$                                                        & $2.94\,(\pm 1.94)$                                                  & $5.43\,(\pm 2.69)$                                                   & $78.68$                             & $85.06$                                      & $78.06$                                              & $6.67$                               \\
               & Noise        & $34.93\,(\pm 2.33)$                                                   & $3.44\,(\pm 2.14)$                                                        & $3.28\,(\pm 2.58)$                                                  & $7.62\,(\pm 2.36)$                                                   & $88.58$                             & $89.03$                                      & $76.82$                                              & $1.74$                               \\ 
    
\thickhline
\end{tabular}
}
\label{tab:1}
\end{table*}

\subsubsection{Methodology Efficacy}

Tab.~\ref{tab:1} presents the performance of the proposed SpikeTimer algorithm on five neuromorphic datasets: N-MNIST, CIFAR10-DVS, DVS128 Gesture, N-Caltech101 and DailyDVS-200. For each dataset, three types of tokens—Static, Moving, and Noise tokens were evaluated across different datasets. 
All experiments will be repeated $8$ times to calculate the mean and standard deviation.

On N-MNIST, SpikeTimer preserves near-baseline performance when the correct token is embedded, achieving authorized accuracy $A_{\text{auth}}$ of $98.96\%$--$99.10\%$ across token types with negligible impact on classification. It strongly rejects unauthorized inputs, reducing both no-token ($A_{\text{no}}$) and mis-tokened ($A_{\text{mis}}$) accuracies to about $8.6\%$--$8.9\%$ ($8.58\%$--$8.91\%$ and $8.42\%$--$8.94\%$, respectively), which yields high effective authentication efficiency $E_{\text{eff}}$ (over $90\%$ for all tokens, specifically $90.77\%$--$91.24\%$). Even under full-token attacks ($A_{\text{full}}$), accuracy stays suppressed at $8.67\%$--$9.34\%$, indicating robustness to brute-force embedding attempts. Temporal robustness is also strong, with $T_{\text{spec}}$ and $T_{\text{cover}}$ exceeding $90\%$ for all token types ($T_{\text{spec}}$: $90.88\%$--$91.26\%$, $T_{\text{cover}}$: $90.34\%$--$91.15\%$). Among tokens, moving and noise tokens slightly outperform static tokens on several metrics. Importantly, the baseline drop $B_{\text{drop}}$ is negligible (max $0.25\%$), showing that token embedding and protection training do not degrade the SNN's core performance on N-MNIST.

CIFAR10-DVS is more challenging due to its higher complexity and lower baseline accuracy ($A_{\text{baseline}} = 81.12\%$). With the correct token, SpikeTimer achieves authorized accuracy $A_{\text{auth}}$ of $80.41\%$--$80.88\%$ across token types, staying close to baseline. It also maintains strong rejection of unauthorized inputs: $A_{\text{no}}$ and $A_{\text{mis}}$ are suppressed to $9.32\%$--$11.71\%$ and $8.39\%$--$12.68\%$, respectively, though moving tokens are slightly more vulnerable. Under full-token attacks ($A_{\text{full}}$), accuracy rises to $12.43\%$--$17.85\%$, indicating moderate but acceptable resistance to brute-force embedding. Performance is token-dependent: static and noise tokens yield higher effectiveness and temporal specificity ($E_{\text{eff}} > 87\%$, $T_{\text{spec}} > 88\%$ for static and noise; moving tokens lag with $E_{\text{eff}}=84.96\%$, $T_{\text{spec}}=83.77\%$), suggesting reduced distinctiveness in this dynamic setting. Timeslice coverage robustness $T_{\text{cover}}$ is lower overall ($77.12\%$--$84.38\%$), again weakest for moving tokens. The baseline drop $B_{\text{drop}}$ remains modest (max $0.88\%$), showing only minor impact on standard performance.

For DVS128 Gesture set, SpikeTimer retains near-baseline performance, with $A_{\text{auth}}$ at $92.59\%$--$93.22\%$ across tokens versus a $93.68\%$ baseline. It strongly rejects unauthorized inputs, reducing $A_{\text{no}}$ to $6.84\%$--$16.28\%$ and $A_{\text{mis}}$ to $6.71\%$--$10.43\%$; the noise token is the most selective (lowest $A_{\text{no}}=6.84\%$ and $A_{\text{mis}}=6.71\%$). Under full-token attacks ($A_{\text{full}}$), static and moving tokens are more vulnerable ($25.41\%$ and $25.86\%$), while the noise token remains resistant ($10.48\%$). Consistently, the noise token achieves the best protection metrics ($E_{\text{eff}}=91.74\%$, $T_{\text{spec}}=91.87\%$, $T_{\text{cover}}=87.85\%$) compared with static ($88.99\%$, $89.25\%$, $72.38\%$) and moving ($81.45\%$, $87.70\%$, $71.24\%$). $B_{\text{drop}}$ is small (max $1.16\%$; $0.96\%$ for noise), indicating that noise tokens provide distinctive temporal signatures with minimal task interference.

Within N-Caltech101, SpikeTimer maintains near-baseline authorized accuracy ($81.86\%$--$83.67\%$) against an $84.23\%$ baseline. Unauthorized inputs are well suppressed, with $A_{\text{no}}$ ranging from $6.54\%$ to $7.89\%$ and $A_{\text{mis}}$ from $7.65\%$ to $8.52\%$. The moving token achieves the highest $E_{\text{eff}}$ ($91.57\%$) and the smallest baseline drop ($0.66\%$), while static and noise tokens also perform reliably ($E_{\text{eff}} > 87\%$). Full-token attack resistance remains strong ($A_{\text{full}} \approx 9.6\%$--$10.9\%$), reflected in $T_{\text{cover}}$ values above $84\%$ for all tokens ($84.21\%$--$87.98\%$). Overall, N-Caltech101 benefits from consistent protection with minimal degradation.

On DailyDVS-200, a challenging dataset with low baseline accuracy ($35.55\%$), SpikeTimer still delivers effective protection. Authorized accuracy reaches $33.18\%$--$34.93\%$, while $A_{\text{no}}$ and $A_{\text{mis}}$ drop to as low as $3.44\%$ and $2.94\%$, respectively. The noise token yields the highest $E_{\text{eff}}$ ($88.58\%$) and a low baseline drop ($1.74\%$). Although $B_{\text{drop}}$ is larger than on other datasets (up to $6.67\%$ for the moving token), all tokens achieve $E_{\text{eff}} > 78\%$ and $T_{\text{spec}} > 85\%$, demonstrating the method's adaptability even under adverse conditions.

\subsubsection{Comparison Analysis}
\label{compare_result}

We conducted comparison between SpikeTimer and comparative methods. Concretely, we include WNB, the first watermarking scheme proposed for SNN copyright protection, and adopt M-LOCK, a classic active copyright-protection method originally designed for general DNNs. 
To further compare SpikeTimer with SOTA backdoor strategies, we additionally incorporated FFCBA and WaveAttack as backdoor baselines.
For evaluation metrics, experiments were performed under three different token settings, and the results were averaged to obtain representative results. Since watermarking is not an active protection mechanism and M-LOCK is not tailored to SNNs, some metrics are not applicable. Therefore, we report $A_{\text{auth}}$, $A_{\text{no}}$, $E_{\text{eff}}$, and $B_{\text{drop}}$.

\begin{table}[t]
\caption{Comparative analysis against prior methods ($\%$).
Please \textit{cf.} Sec. \ref{compare_result} for detailed explanations.}
\label{exp-compare}
\centering
\resizebox{1\columnwidth}{!}{
\renewcommand{\arraystretch}{1.}
\begin{tabular}{l|llll}
\thickhline
\multicolumn{1}{c|}{\textbf{Token Type}} & \multicolumn{1}{c}{\textbf{$A_{\text{auth}}$}} & \multicolumn{1}{c}{\textbf{$A_{\text{no}}$}} & \multicolumn{1}{c}{\textbf{$E_{\text{eff}}$}} & \multicolumn{1}{c}{\textbf{$B_{\text{drop}}$}} \\ 
\hline
\hline
\multicolumn{5}{l}{\textcolor{gray!60}{\textit{Implement in N-MNIST Dataset}}} \\
WNB                         & $98.35$                & $98.35$                & $0.00$                 & $0.90$                \\
M-LOCK                      & $98.66$                & $9.79$                 & $88.87$                & $0.58$                 \\
FFCBA                       & $97.83$                & $11.06$                & $87.44$                & $1.42$                 \\
WaveAttack                  & $97.41$                & $12.58$                & $85.48$                & $1.84$                 \\
\hline
SpikeTimer                  & $99.16$                & $8.79$                 & $90.37$                & $0.08$                \\ 
\hline
\hline
\multicolumn{5}{l}{\textcolor{gray!60}{\textit{Implement in CIFAR10-DVS Dataset}}} \\
WNB                         & $79.23$                & $78.23$                & $0.00$                 & $2.24$                \\
M-LOCK                      & $78.45$                & $11.21$                & $82.95$                & $3.21$                 \\
FFCBA                       & $77.69$                & $13.28$                & $79.47$                & $4.15$                 \\
WaveAttack                  & $76.03$                & $14.86$                & $75.47$                & $6.19$                 \\
\hline
SpikeTimer                  & $80.66$                & $10.21$                & $86.90$                & $0.49$                \\ 
\hline
\hline
\multicolumn{5}{l}{\textcolor{gray!60}{\textit{Implement in DVS128 Gesture Dataset}}} \\
WNB                         & $90.42$                & $90.42$                & $0.00$                 & $3.50$                \\
M-LOCK                      & $92.75$                & $15.31$                & $82.65$                & $1.01$                 \\
FFCBA                       & $91.80$                & $17.23$                & $79.58$                & $2.03$                 \\
WaveAttack                  & $90.04$                & $19.71$                & $75.06$                & $3.91$                 \\
\hline
SpikeTimer                  & $93.25$                & $11.06$                & $87.72$                & $0.48$                \\ 
\thickhline
\end{tabular}%
}
\end{table}

Tab.~\ref{exp-compare} reports a quantitative comparison between SpikeTimer and four baselines: WNB (passive watermarking), M-LOCK (active protection for DNNs), and two SOTA backdoor strategies FFCBA and WaveAttack. Overall, SpikeTimer consistently achieves the most favorable protection. It maintains high accuracy on authorized tokenized inputs ($A_{\text{auth}}$), enforces low accuracy on non-token inputs ($A_{\text{no}}$), obtains the highest protection effectiveness ($E_{\text{eff}}$), and introduces the smallest benign performance degradation ($B_{\text{drop}}$).

On N-MNIST, SpikeTimer reaches $A_{\text{auth}}=99.16\%$ and suppresses non-token inputs to $A_{\text{no}}=8.79\%$, yielding $E_{\text{eff}}=90.37\%$ with a negligible $B_{\text{drop}}=0.08\%$. Among the baselines, M-LOCK achieves $A_{\text{auth}}=98.66\%$ and $A_{\text{no}}=9.79\%$, but its $E_{\text{eff}}=88.87\%$ and $B_{\text{drop}}=0.58\%$ are both less favorable. FFCBA and WaveAttack, though designed as backdoor attacks, show lower authorized accuracy ($97.83\%$ and $97.41\%$, respectively) and higher non-token accuracy ($11.06\%$ and $12.58\%$), resulting in $E_{\text{eff}}=87.44\%$ and $85.48\%$ with larger $B_{\text{drop}}$ ($1.42\%$ and $1.84\%$). FFCBA and WaveAttack are originally designed for static image backdoor attacks and do not accommodate the event‑driven dynamics of SNNs, and consequently exhibit weaker suppression of unauthorized inputs and higher efficacy degradation compared to SpikeTimer. WNB, as a passive watermarking scheme, provides no active protection ($E_{\text{eff}}=0.00\%$, $A_{\text{no}}=98.35\%$) and also suffers a $B_{\text{drop}}$ of $0.90\%$.

For CIFAR10-DVS, SpikeTimer again delivers the best overall results: $A_{\text{auth}}=80.66\%$, $A_{\text{no}}=10.21\%$, $E_{\text{eff}}=86.90\%$, and $B_{\text{drop}}=0.49\%$. M-LOCK is less effective ($E_{\text{eff}}=82.95\%$) and incurs a much larger benign degradation ($B_{\text{drop}}=3.21\%$). FFCBA and WaveAttack show further reduced $A_{\text{auth}}$ ($77.69\%$ and $76.03\%$) and elevated $A_{\text{no}}$ ($13.28\%$ and $14.86\%$), leading to lower $E_{\text{eff}}$ ($79.47\%$ and $75.47\%$) and higher $B_{\text{drop}}$ ($4.15\%$ and $6.19\%$). WNB remains ineffective, with $A_{\text{no}}=78.23\%$ nearly matching its $A_{\text{auth}}=79.23\%$, yielding $E_{\text{eff}}=0.00\%$.

Regarding DVS128 Gesture, SpikeTimer maintains strong performance: $A_{\text{auth}}=93.25\%$, $A_{\text{no}}=11.06\%$, $E_{\text{eff}}=87.72\%$, and $B_{\text{drop}}=0.48\%$. M-LOCK achieves $E_{\text{eff}}=82.65\%$ and $B_{\text{drop}}=1.01\%$. FFCBA and WaveAttack yield $E_{\text{eff}}=79.58\%$ and $75.06\%$, respectively, with larger $B_{\text{drop}}$ ($2.03\%$ and $3.91\%$). WNB provides no effective protection ($E_{\text{eff}}=0.00\%$, $A_{\text{no}}=90.42\%$), which reflects its passive nature.

\subsubsection{Computational Overhead Analysis}
\label{overhead_result}
To evaluate the computational and deployment overhead of SpikeTimer, we further compare it with the original SNN on N-MNIST, CIFAR10-DVS, and DVS128 Gesture. To keep a fair comparison, we made all model-side and system-side factors unchanged (\textit{i.e.}, the network architecture, number of timesteps, optimizer, batch size, and hardware environment). Consequently, the parameter count of the protected model is identical to that of the original SNN by design. We report the training time and inference latency under this controlled setting.
Concerning inference latency, the original SNN was evaluated on clean test samples, whereas SpikeTimer was tested on authorized samples with the valid token embedded in the designated timeslice, since this corresponds to the normal deployment scenario of legitimate users.
The empirical results are listed in Tab.~\ref{exp-overhead}.

\begin{table}[t]
\caption{Computational overhead comparison against prior methods (training time: \textit{s/epoch}; inference latency: \textit{ms/sample}).
Please \textit{cf.} Sec. \ref{overhead_result} for detailed explanations.}
\label{exp-overhead}
\centering
\resizebox{0.98\columnwidth}{!}{
\renewcommand{\arraystretch}{1.2}
\begin{tabular}{l|p{0.14\columnwidth}p{0.14\columnwidth}p{0.14\columnwidth}}
\thickhline
\multicolumn{1}{c|}{\textbf{Metric}} & 
\multicolumn{1}{>{\centering\arraybackslash}p{0.14\columnwidth}}{\textbf{WNB}} & 
\multicolumn{1}{>{\centering\arraybackslash}p{0.14\columnwidth}}{\textbf{M-LOCK}} & 
\multicolumn{1}{>{\centering\arraybackslash}p{0.14\columnwidth}}{\textbf{SpikeTimer}} \\ 
\hline
\hline
\multicolumn{4}{l}{\textcolor{gray!60}{\textit{Implement in N-MNIST Dataset}}} \\
Training Time                         & \centering $45.67$                & \centering $58.90$                & \centering\arraybackslash $49.12$                 \\
Inference Latency                     & \centering $1.04$                 & \centering $3.58$                 & \centering\arraybackslash $2.27$                 \\
\hline
\hline
\multicolumn{4}{l}{\textcolor{gray!60}{\textit{Implement in CIFAR10-DVS Dataset}}} \\
Training Time                         & \centering $180.25$               & \centering $265.33$               & \centering\arraybackslash $192.81$                \\
Inference Latency                     & \centering $3.20$                 & \centering $7.41$                 & \centering\arraybackslash $5.79$                 \\
\hline
\hline
\multicolumn{4}{l}{\textcolor{gray!60}{\textit{Implement in DVS128 Gesture Dataset}}} \\
Training Time                         & \centering $58.50$                & \centering $72.94$                & \centering\arraybackslash $64.16$                 \\
Inference Latency                     & \centering $2.85$                 & \centering $6.51$                 & \centering\arraybackslash $4.08$                 \\
\thickhline
\end{tabular}%
}
\end{table}

As evidenced in Tab.~\ref{exp-overhead}, WNB achieves the lowest computational cost because it is a passive watermarking scheme and does not require an additional generator to synthesize authorized samples. In contrast, both M-LOCK and SpikeTimer involve generator training to construct token-embedded authorized samples, which inevitably increases the training cost. Nevertheless, the additional overhead of SpikeTimer remains limited compared with WNB. Specifically, the training time of SpikeTimer is $49.12~\mathrm{s/epoch}$, $192.81~\mathrm{s/epoch}$, and $64.16~\mathrm{s/epoch}$ on N-MNIST, CIFAR10-DVS, and DVS128 Gesture, respectively, corresponding to only $7.55\%$, $6.97\%$, and $9.68\%$ increases over WNB. For inference, SpikeTimer reports latencies of $2.27~\mathrm{ms/sample}$, $5.79~\mathrm{ms/sample}$, and $4.08~\mathrm{ms/sample}$, with absolute increases of only $1.23~\mathrm{ms/sample}$, $2.59~\mathrm{ms/sample}$, and $1.23~\mathrm{ms/sample}$ over WNB. These moderate gaps indicate that SpikeTimer achieves comparable efficiency to the lightweight passive watermarking method, while providing stronger protection under illegal model-copying scenarios where WNB lacks active robustness. Moreover, compared with M-LOCK, SpikeTimer reduces the training time by $16.60\%$, $27.33\%$, and $12.04\%$, and decreases the inference latency by $36.59\%$, $21.86\%$, and $37.33\%$ on the three datasets, respectively. These results demonstrate that SpikeTimer offers a favorable trade-off between computational efficiency and active protection capability.

\subsection{Ablation Study}
We evaluated the parameter impact. Unless otherwise specified, we follow the main experiment settings described before and implement all experiments using SpikingJelly \cite{adi1480}.
Note that, for SpikingJelly, backbone CIFAR10-DVS accuracy $A_{\text{baseline}}= 68.03\%$.

\subsubsection{Duration of Authorized Timeslice}
\label{timeslice duration}
\begin{figure}[bt]
  \centering
\includegraphics[width=\linewidth]{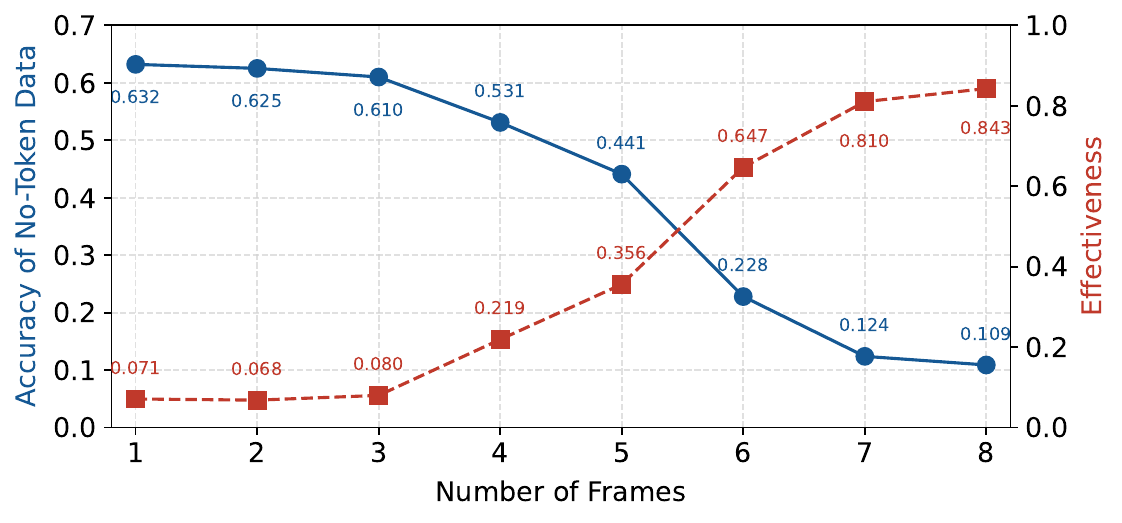}
  \caption{Duration of authorized timeslice experiment on CIFAR10-DVS with static token. Please \textit{cf.} Sec. \ref{timeslice duration} for detailed explanations.}
  \label{fig:8}
\end{figure}

We examined how the duration of the authorized timeslice affects system performance. For test setting, each neuromorphic sample contains $16$ frames and was evenly partitioned into an authorized segment and an unauthorized segment, where the first $8$ frames are treated as the authorized timeslice and the remaining $8$ frames as the unauthorized timeslice. To quantify the impact of shortening the authorized duration, we kept the total number of frames fixed and progressively reduced the number of authorized frames from $8$ to $1$, assigning all remaining frames to the unauthorized timeslice. This controlled design enables a systematic evaluation of the resulting changes in accuracy and robustness, and further reveals the trade-offs among embedding capacity, detectability, and protection effectiveness. The study also highlights the flexibility and practical limitations of SpikeTimer in settings where only short authorized timeslices are feasible due to data constraints or operational requirements.
The results are summarized in Fig.~\ref{fig:8}.

On CIFAR10-DVS with the static Token, authorized performance remains stable across all configurations: $A_{\text{auth}}$ consistently stays at $66.4\%$ or higher, only slightly below the baseline without SpikeTimer, indicating that varying the authorized timeslice length does not compromise authorized usability. In contrast, the no-token (unauthorized) accuracy exhibits a pronounced dependence on the authorized duration, revealing a clear protection trade-off. With only one authorized frame, $A_{\text{no}}$ remains high at $63.2\%$, whereas extending the authorized timeslice progressively suppresses unauthorized inference, reducing $A_{\text{no}}$ to $10.9\%$ when $8$ frames are authorized. Consistently, the effectiveness metric increases as the authorized timeslice length grows, reaching $84.27\%$ at $8$ frames, which confirms that longer authorized timeslices substantially strengthen SpikeTimer’s ability to disrupt unauthorized usage.

\subsubsection{Data Distribution Influence}
\label{data distribute}
We conducted experiments by varying the distribution proportions among the Authorized Dataset, the Full-Token Dataset, and the No-Token Dataset. For configuration, the data distribution was set to $[0.85, 0.1, 0.05]$, where $85\%$ of the data belonged to the authorized dataset, $10\%$ to the full-token dataset, and $5\%$ to the no-token dataset (contrastive loss is neglected).



\begin{figure}[t]
    \centering
    \subcaptionbox{Full-Token dataset\protect\label{fig:6(a)}}
      {\includegraphics[width=0.48\linewidth]{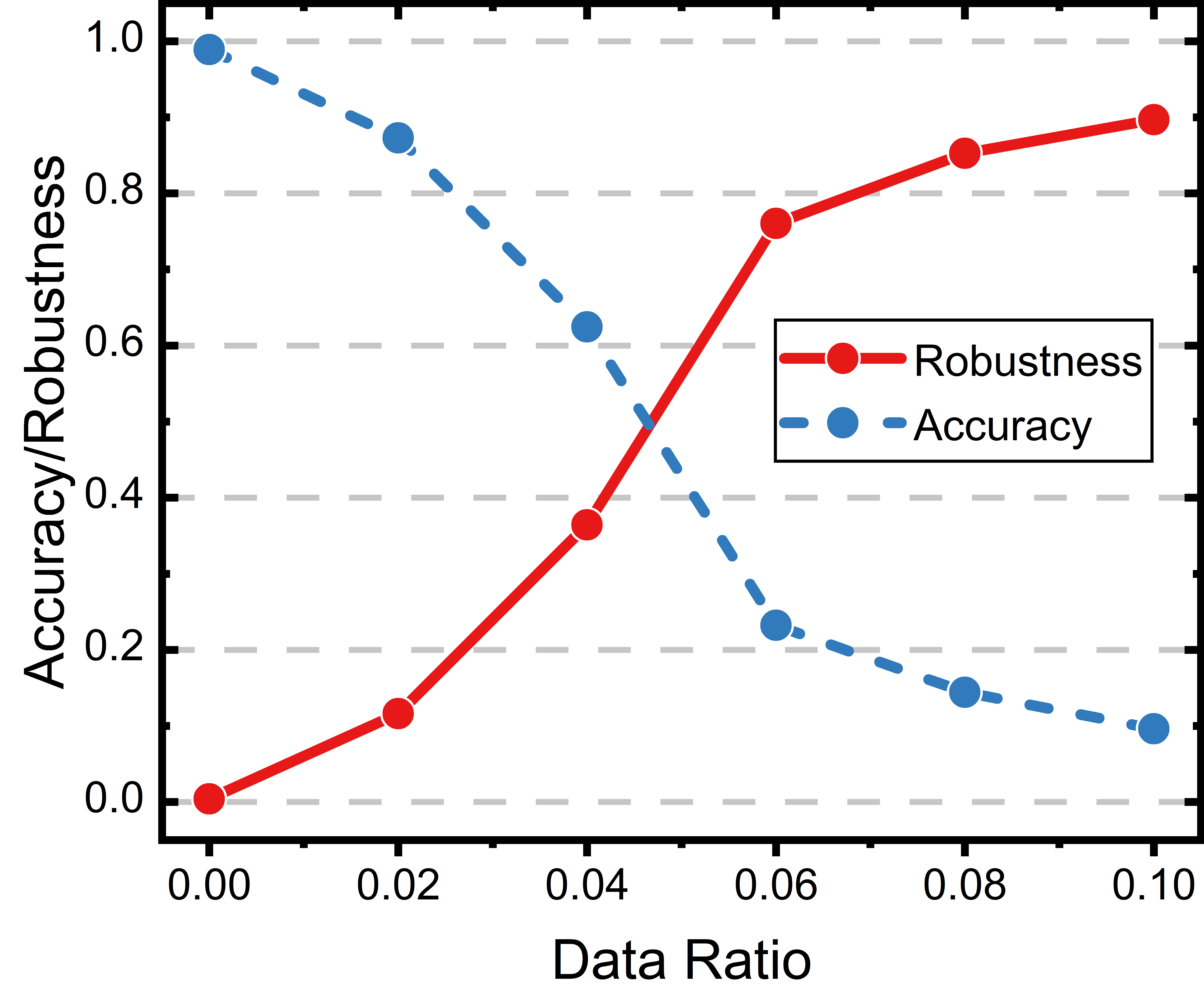}}
    \subcaptionbox{No-Token dataset\protect\label{fig:6(b)}}
      {\includegraphics[width=0.48\linewidth]{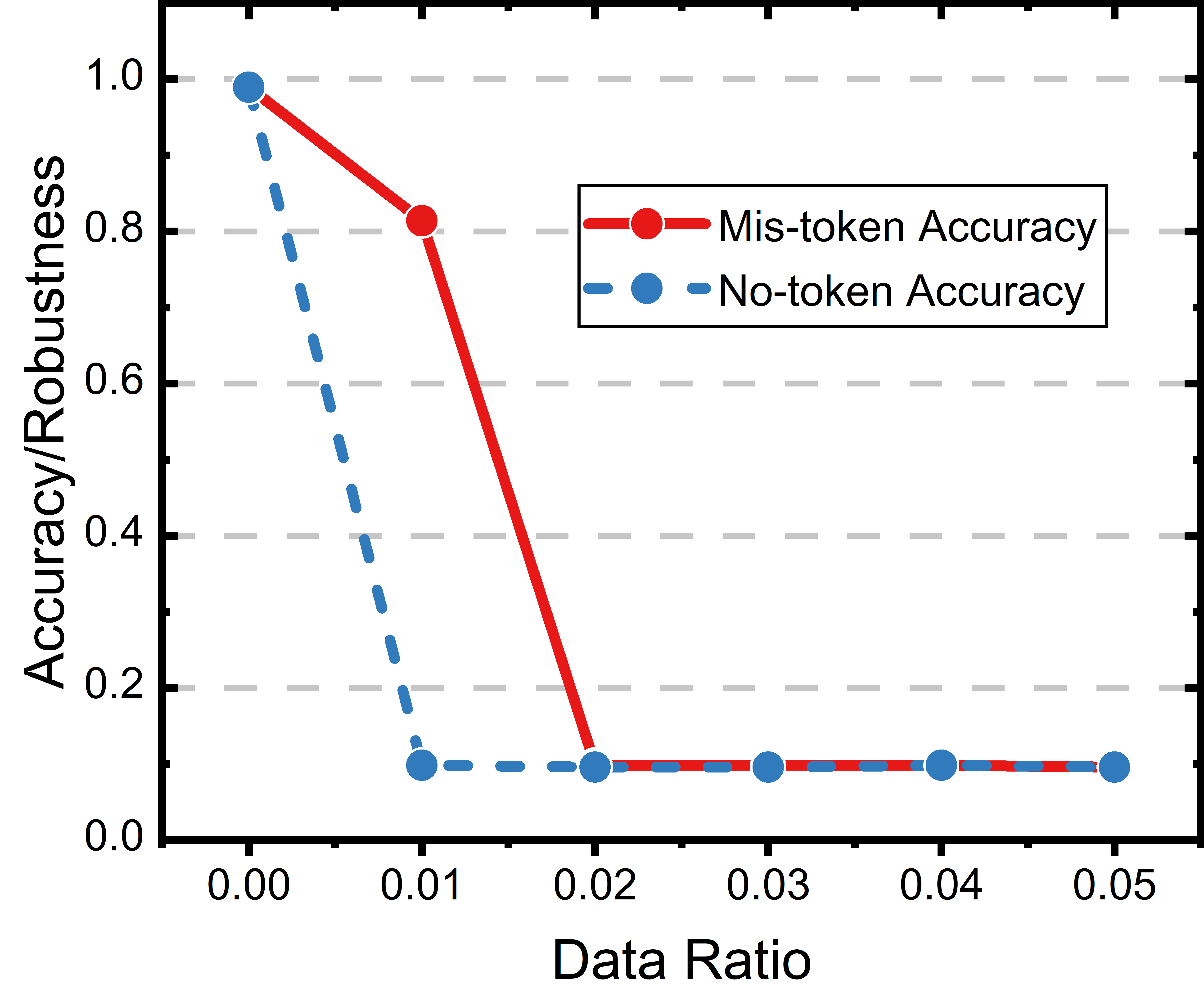}}
    \caption{Training data distribution experiment on N-MNIST with full-token dataset and no-token dataset. Please \textit{cf.} Sec. \ref{data distribute} for detailed explanations.}
    \label{fig:9}
\end{figure}

The experimental results on the N-MNIST dataset are shown in Fig. \ref{fig:9}.
In Fig. \ref{fig:6(a)}, the proportion of the no-token Dataset is fixed at $5\%$, while the ratio of the full-token set is incrementally adjusted from $0\%$ to $10\%$.  It is important to note that full-token data represents an exhaustive token injection strategy, akin to a brute-force attack, where tokens are injected across all possible timeslices, violating the mechanism’s requirement that authorized data tokens must be confined to specific timeslices.
This experiment examines the effect of full-token data on both full-token accuracy and Timeslice Coverage Robustness. When the full-token ratio is zero, the accuracy remains close to $100\%$, as expected. As the ratio increases, the accuracy drops sharply and approaches approximately $10\%$ at the maximum setting, indicating effective resistance to brute-force token injection. Lower full-token accuracy reflects the system’s ability to detect unauthorized inputs and prevent circumvention of temporal constraints. Meanwhile, the Timeslice Coverage Robustness ($T_{\text{cover}}$) increases monotonically with the full-token ratio, reaching a maximum of $0.8975$, demonstrating that SpikeTimer increasingly distinguishes legitimate authorized data from adversarial full-token inputs as attack intensity grows. Overall, these results confirm the robustness of the framework against unauthorized temporal embeddings.


In Fig.~\ref{fig:6(b)}, we fix the full-token ratio at $10\%$ and progressively increase the rate of the no-token dataset from $0\%$ to $5\%$, where no-token samples contain no temporal tokens and thus simulate unauthorized inputs. As the rate rises, the no-token accuracy drops sharply from nearly $100\%$ to approximately $0\%$, indicating that even a small fraction of samples can substantially impair the model ability to produce reliable predictions on unauthorized data. In parallel, the mis-token accuracy $A_{\text{mis}}$ also decreases to near zero, suggesting that SpikeTimer consistently suppresses unauthorized inference across different temporal embedding patterns. Overall, these results show that SpikeTimer effectively counters both brute-force full-token injection and unauthorized no-token access by driving adversarial accuracies toward zero while preserving temporal authorization sensitivity.



\subsubsection{Token Locations}
\label{token-location}
We evaluated the impact of the token’s spatial placement. Beyond the default configuration in main experiments, the trigger location is restricted to a predefined set of candidate regions: \{\textit{top-left, top-right, bottom-left, bottom-right, middle}\}. Evaluation is conducted on MSIL model and CIFAR10-DVS sets via static token patterns.
The empirical results are listed in Tab.~\ref{exp-token-location}.

\begin{table}[t]
\caption{SpikeTimer performances ($\%$) across token locations with MSIL architecture. Please \textit{cf.} Sec. \ref{token-location} for detailed explanations.}
\label{exp-token-location}
\centering
\resizebox{0.90\columnwidth}{!}{
\renewcommand{\arraystretch}{1.2}
\begin{tabular}{l|llll}
\thickhline
\multicolumn{1}{c|}{\textbf{Token Location}} 
& \multicolumn{1}{c}{\textbf{$A_{\text{auth}}$}} 
& \multicolumn{1}{c}{\textbf{$A_{\text{no}}$}} 
& \multicolumn{1}{c}{\textbf{$A_{\text{mis}}$}} 
& \multicolumn{1}{c}{\textbf{$A_{\text{full}}$}} \\ 
\hline\hline
\textit{Top-Left}
& $81.25$  & $8.46$  & $9.37$  & $19.83$  \\
\textit{Top-Right} 
& $81.91$  & $9.23$ & $10.50$ & $20.41$  \\
\textit{Bottom-Left}  
& $82.08$  & $8.14$  & $8.95$  & $18.67$  \\ 
\textit{Bottom-Right}   
& $80.52$  & $9.77$  & $9.88$  & $19.25$  \\ 
\textit{Middle}
& $81.96$  & $8.05$  & $8.82$  & $18.30$  \\ 
\thickhline
\end{tabular}%
}
\end{table}

Overall, the results are highly consistent across all five placements, indicating that the token location has negligible influence on the defense behavior. In particular, $A_{\text{auth}}$ remains stable within a narrow range of $80.52\%\sim 82.08\%$, showing that legitimate authentication accuracy is largely invariant to spatial placement. Meanwhile, the reversed/no-token accuracies stay uniformly low: $A_{\text{no}}$ varies only from $8.05\%\sim 9.77\%$ and $A_{\text{mis}}$ ranges from $8.82\%\sim 10.50\%$, suggesting that incorrect or absent tokens remain ineffective regardless of where the trigger is placed. The overall metric $A_{\text{full}}$ is similarly stable, spanning $18.30\%\sim 20.41\%$. 
Although the spatial location of the token is not the core novelty of SpikeTimer, this ablation is included to verify that the proposed temporal authorization mechanism does not depend on a specific trigger placement.
These results collectively confirm that SpikeTimer is robust to token spatial placement, and relocating the token within the predefined candidate regions does not materially affect the system performance.

\subsection{Robustness Study}
To assess SpikeTimer’s security, we conducted a robustness analysis against adversarial threats via SpikingJelly.
The empirical analysis is provided in Sec.~\ref{sec_analyze}.

\subsubsection{Model Fine-Tuning Attacks}
\label{exp-ft-atk}



The attack was implemented according to \cite{guo2019spottune}.
We assessed this using cross-dataset training. For the CIFAR10-DVS model, we performed fine-tuning by using the DVS128 Gesture dataset, while for the DVS128 Gesture model, we fine-tuned it using the CIFAR10-DVS dataset. It is important to note that if an attacker can obtain private training data, they could simply train a new model from scratch, bypassing the demand to fine-tune a protected model.

\begin{figure}[t]
    \centering
    \subcaptionbox{CIFAR10-DVS model\protect\label{fig:10a}}
      {\includegraphics[width=0.49\linewidth]{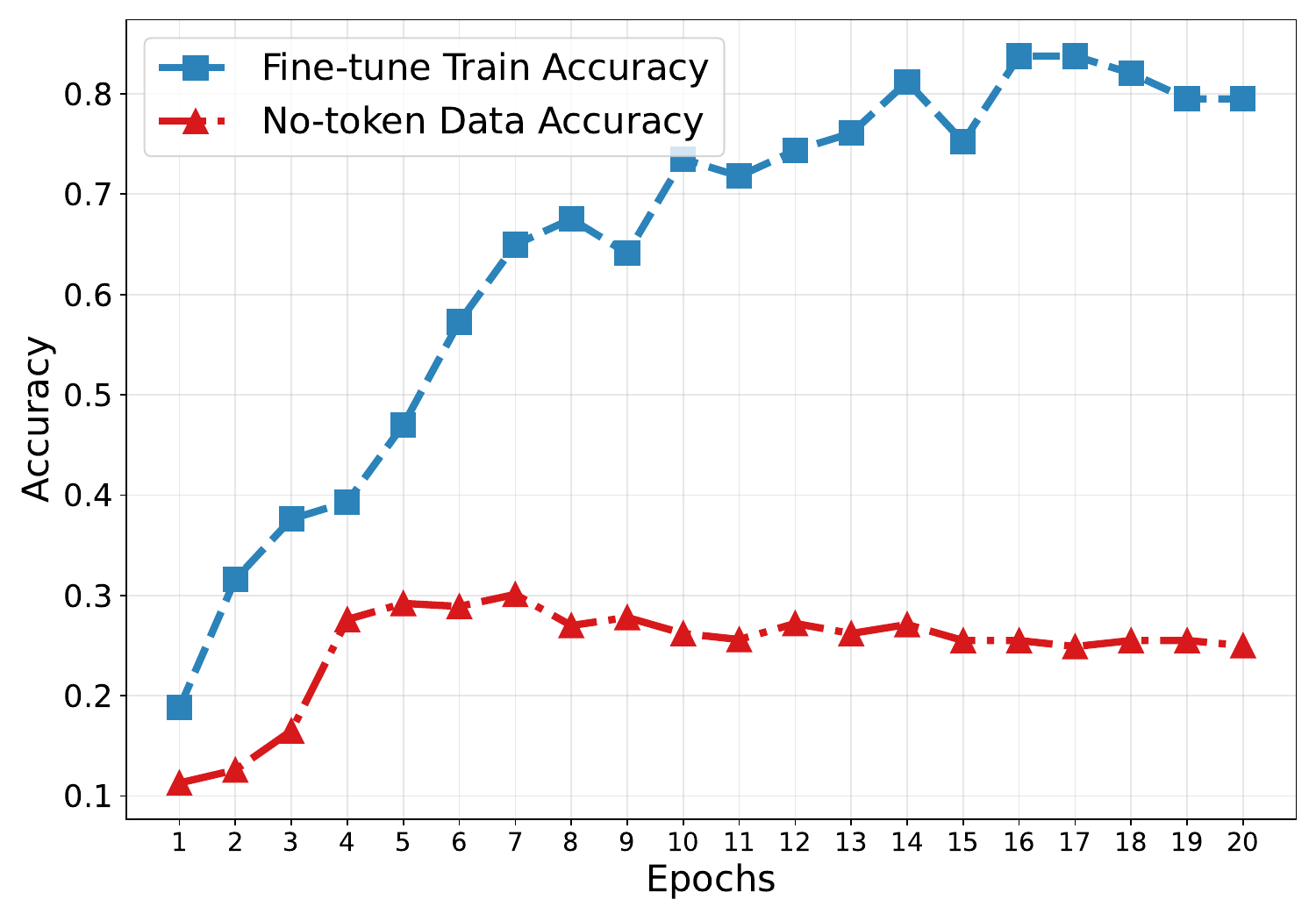}}
    \subcaptionbox{DVS128 gesture model\protect\label{fig:10b}}
      {\includegraphics[width=0.49\linewidth]{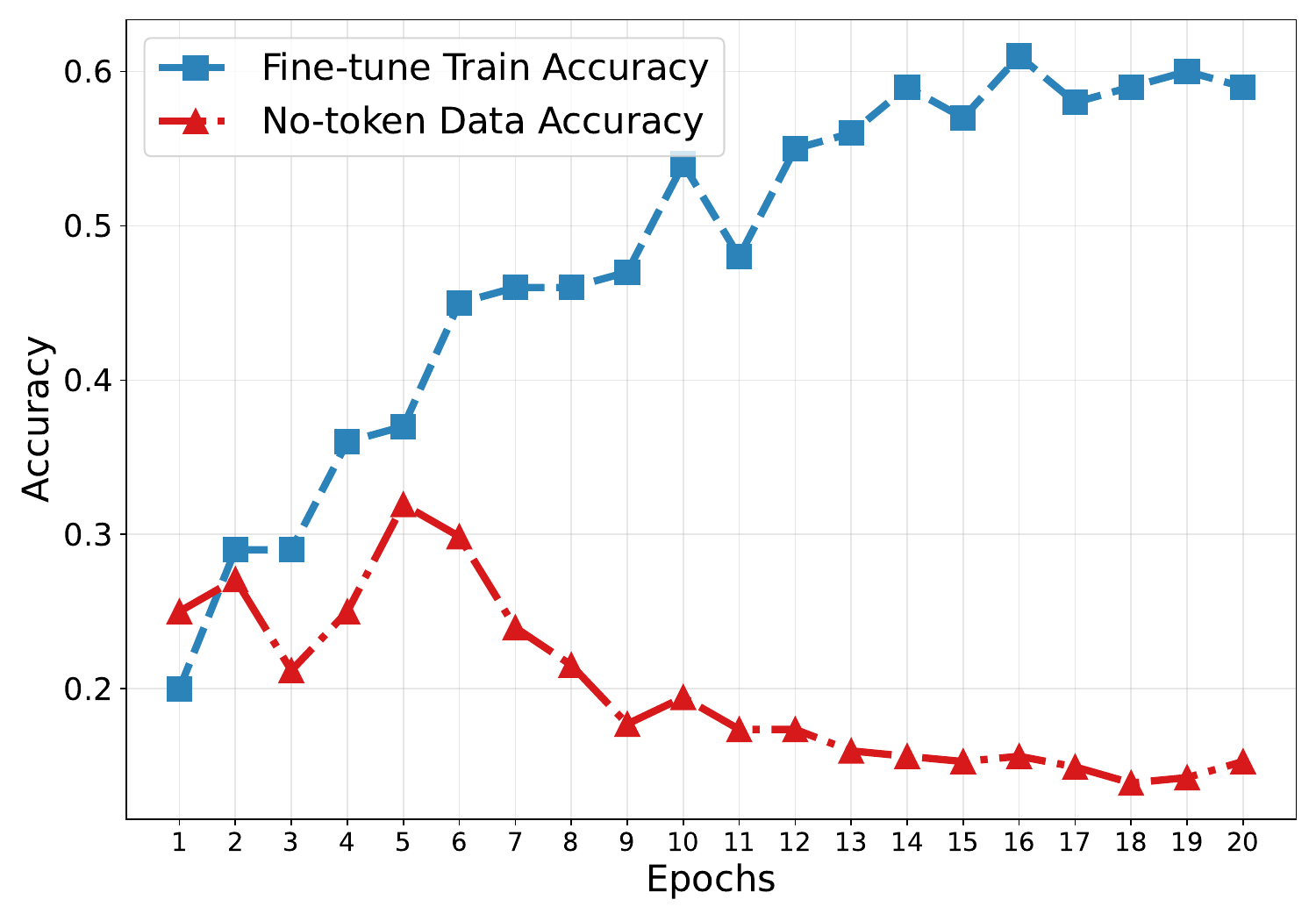}}
    \caption{SpikeTimer models' performances under \textit{model fine-tuning} attacks. Please \textit{cf.} Sec. \ref{exp-ft-atk} for detailed explanations.}
    \label{fig:10}
\end{figure}

The evaluation outcomes are provided in Fig.~\ref{fig:10}.
For both the CIFAR10-DVS model (Fig. \ref{fig:10a}) and the DVS128 Gesture model (Fig. \ref{fig:10b}), we observe the following trends: The accuracy of the fine-tuned model on the training data steadily improves over the training epochs, reaching high values by the end of the fine-tuning process. This indicates that the model adapts well to the new data domain introduced during fine-tuning. Despite the improvement in fine-tune train accuracy, the accuracy on no-token data remains significantly lower and does not exhibit substantial growth over the epochs. This demonstrates that the SpikeTimer mechanism effectively maintains robustness against unauthorized data usage, even when the model is subjected to cross-domain fine-tuning.


\subsubsection{Model Pruning Attacks}
\label{exp-prune-atk}

\begin{figure}[t]
    \centering
    \subcaptionbox{Static token\protect\label{fig:11a}}
      {\includegraphics[width=\linewidth]{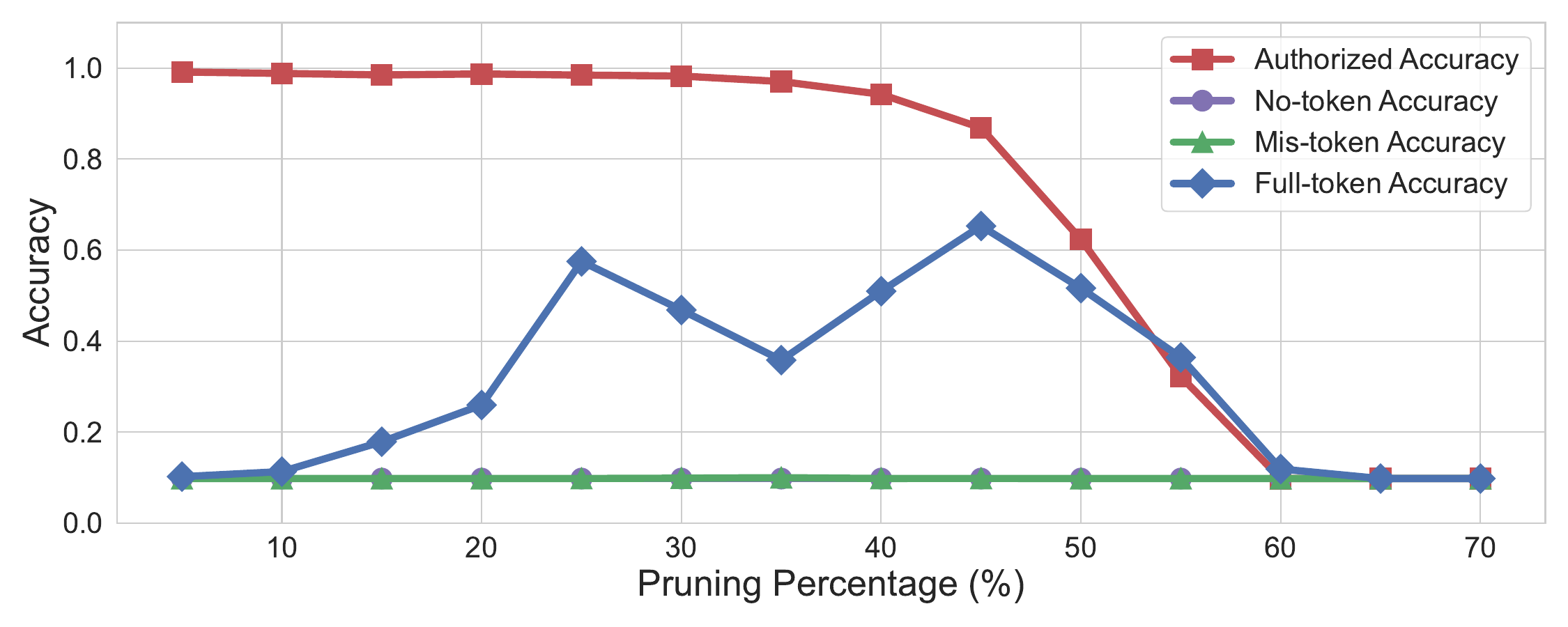}}
    \subcaptionbox{Moving token\protect\label{fig:11b}}{\includegraphics[width=\linewidth]{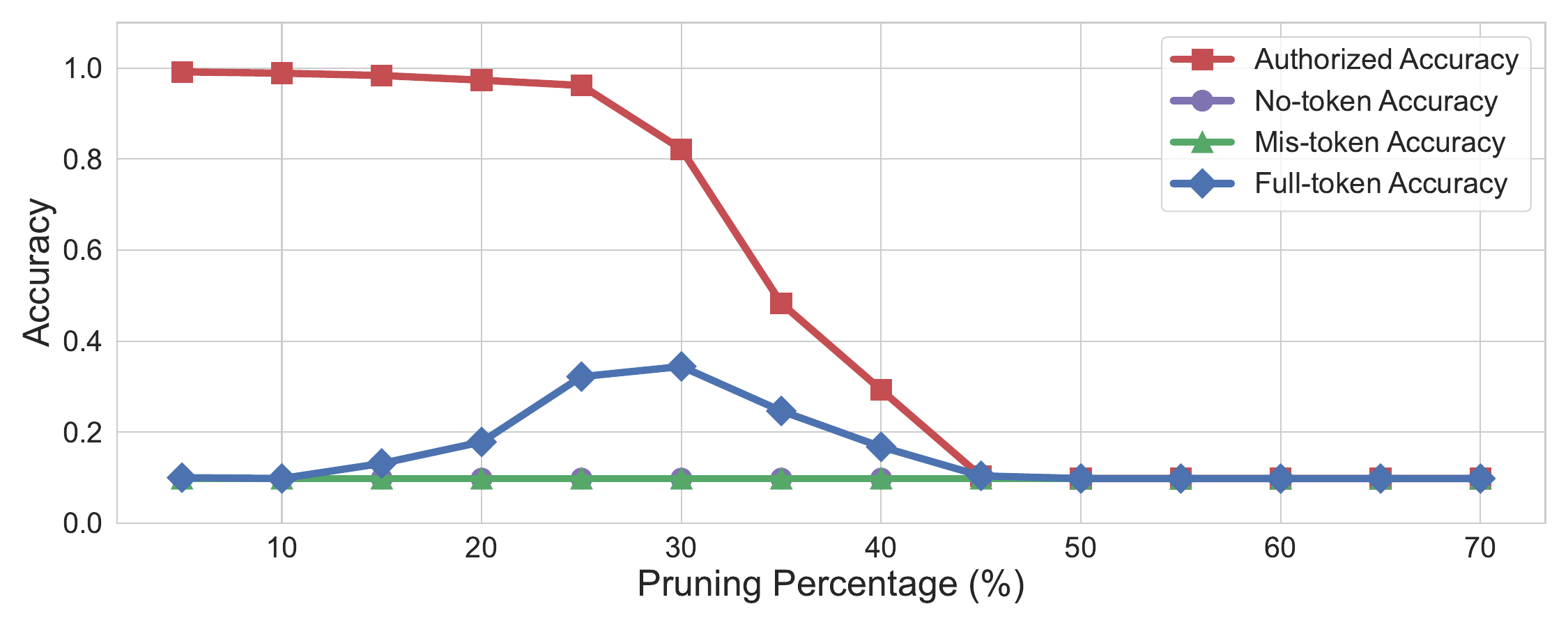}}
    \subcaptionbox{Noise token\protect\label{fig:11c}}{\includegraphics[width=\linewidth]{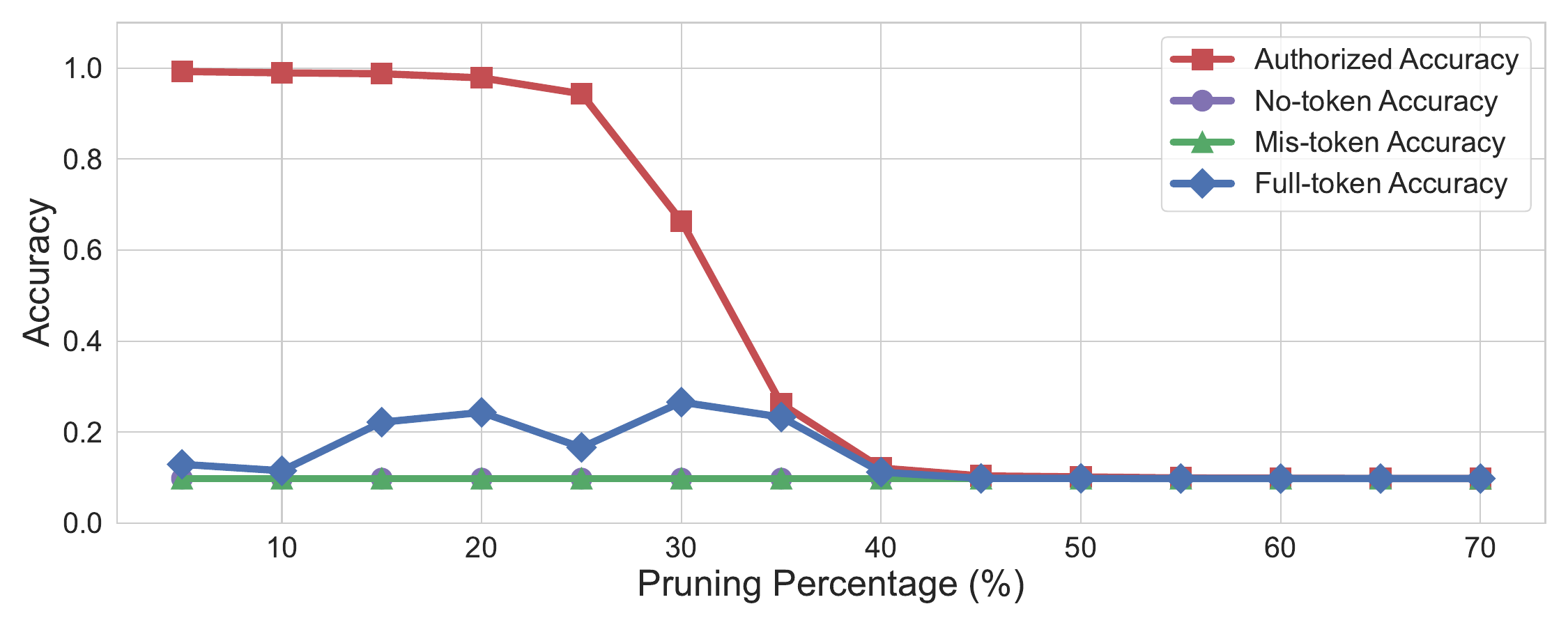}}
    
    \caption{SpikeTimer performances under \textit{model pruning} on N-MNIST. Please \textit{cf.} Sec. \ref{exp-prune-atk} for detailed explanations.
    }
    \label{exp-prune}
\end{figure}

We evaluated this robustness under different pruning percentages with the method described in \cite{liu2021discrimination}. 
The results are illustrated in Fig.~\ref{exp-prune}.



From the outcomes, in all scenarios (Static, Moving, and Noise Token), the authorized accuracy remains high and stable until a significant pruning level (around $30\%-50\%$), highlighting the model's ability to maintain functionality when used with authorized tokens. Both no-token and mis-token accuracy remain consistently low across all pruning percentages, confirming that unauthorized access is effectively prevented and unaffected by pruning. Although the full-token accuracy shows a brief increase at intermediate pruning levels, it remains significantly lower than the authorized accuracy and never reaches a usable level. This indicates that the SpikeTimer mechanism effectively thwarts brute-force token embedding attacks, even under pruning.
Compared to other types of tokens, noise tokens pose greater difficulty in achieving improvements in full-token accuracy after pruning. This could be attributed to the inherent randomness of noise, which diminishes its alignment with the model's features.


\subsubsection{Reverse-Engineering Attacks}
\label{exp-reverse-eng-atk}

We applied the method in \cite{8835365} to perform real-time reverse engineering, reconstructing a token for each timeslice, then injecting these tokens into the overall spiking input stream. Accuracy metrics for reversed data was subsequently monitored to assess whether the method can still maintain its original defensive efficacy. We used SpikingJelly architecture for evaluation.
Tab.~\ref{tab-reverse-atk} summarizes SpikeTimer’s performance.

\begin{table}[t]
\centering
\caption{SpikeTimer performances ($\%$) under \textit{Reverse-Engineering}. Please \textit{cf.} Sec. \ref{exp-reverse-eng-atk} for detailed explanations.}
\label{tab-reverse-atk}
\resizebox{0.95\columnwidth}{!}{
\renewcommand{\arraystretch}{1.1}
\begin{tabular}{l|ccc}
\thickhline
\multicolumn{1}{c|}{\textbf{Token Type}} & \multicolumn{1}{c}{N-MNIST} & \multicolumn{1}{c}{CIFAR10-DVS } & \multicolumn{1}{c}{DVS128 Gesture} \\ 
\hline
\hline
\multicolumn{4}{l}{\textcolor{gray!60}{\textit{Implemented on the CIFAR10-DVS Dataset}}} \\
Static  
& $32.04$  
& $19.82$  
& $21.50$  \\
Moving  
& $9.70$  
& $10.45$  
& $13.11$  \\
Noise   
& $9.20$  
& $8.93$  
& $11.50$      \\ \thickhline
\end{tabular}%
}
\end{table}

The results are unequivocal: across all datasets and token types, the reversed spike streams yield uniformly poor recognition accuracy. This substantiates that the attacker fails to recover discriminative information from the protected input, and SpikeTimer’s defensive efficacy is preserved.
Among all settings, static tokens show the highest (yet still inadequate) accuracies $32.04\%$ on N-MNIST and below $22\%$ on CIFAR10-DVS and DVS128 Gesture. This is expected, since static patterns are structurally simpler, making them comparatively easier to approximate. 
More importantly, moving and noise tokens are effectively non-recoverable: accuracies remain consistently below $14\%$ on all datasets. Moving tokens encode strong spatiotemporal dependencies. Any reconstruction error quickly destroys the temporal coherence required for correct classification. Noise tokens further exacerbate the difficulty by enforcing discrete and stochastic event distributions, rendering reverse-engineered sequences largely random and non-informative. Consequently, the reverse-engineering pipeline collapses into producing spike streams that are functionally useless for downstream recognition.



\section{Discussion}
\label{sect6}

\subsection{Generalization across Model Architectures}
\label{general-model}

To further examine the architectural generalization of SpikeTimer for proactive copyright protection, we incorporated recent architectural advancements from SNNs: 1) Spikingformer \cite{DBLP:conf/aaai/ZhouYZZWZMT26}, 2) SD-Transformer \cite{NEURIPS2023_ca0f5358}, 3) MSIL \cite{DBLP:conf/aaai/ShanZ0QEQ25}, and 4) TE-Spikformer \cite{GAO2024128268}. Spikingformer addresses non-spike computations in spiking transformers via spike-driven residual self-attention. 
SD-Transformer features event-driven binary spike communication, linear-complexity self-attention based solely on mask and addition operations, which holds ultra-low-energy SNN inference.
MSIL introduces Spiking Multiscale Attention to capture multiscale spatiotemporal interactions in event streams and incorporates Attention ZoneOut as a ZoneOut-based regularization strategy to power pseudo-ensemble training and improve generalization. TE-Spikformer adopts Batch Group Normalization to alleviate the imbalance of average firing rates across temporal steps, thereby stabilizing temporal representations. It further enhances temporal information modeling in event-based data by spiking spatiotemporal attention mechanism.
The empirical outcomes are detailed in Tab.~\ref{exp-general}.

\begin{table}[t]
\caption{SpikeTimer performances ($\%$) on CIFAR10-DVS with different SNN architectures. Please \textit{cf.} Sec. \ref{general-model} for detailed explanations.}
\label{exp-general}
\resizebox{\columnwidth}{!}{%
\renewcommand{\arraystretch}{1.2}
\begin{tabular}{l|l|llll}
\thickhline
\multicolumn{1}{c|}{\textbf{Architecture}} & \multicolumn{1}{c|}{\textbf{Token Type}} & \multicolumn{1}{c}{\textbf{$A_{\text{auth}}$}} & \multicolumn{1}{c}{\textbf{$A_{\text{no}}$}} & \multicolumn{1}{c}{\textbf{$A_{\text{mis}}$}} & \multicolumn{1}{c}{\textbf{$A_{\text{full}}$}} \\ \hline\hline
& \multicolumn{5}{l}{{\textit{Backbone Accuracy $A_{\text{baseline}}= 81.49\%$}}} \\
\multirow{3}{*}{Spikingformer}
    & Static   & $80.23$ & $11.27$ & $10.13$ & $14.32$ \\
    & Moving   & $79.88$ & $10.56$ & $9.62$  & $18.67$ \\
    & Noise    & $80.47$ & $11.93$ & $10.85$ & $13.05$ \\ \hline\hline
\multirow{3}{*}{SD-Transformer}
    & \multicolumn{5}{l}{{\textit{Backbone Accuracy $A_{\text{baseline}}= 76.01\%$}}} \\
    & Static   & $74.63$ & $9.45$  & $11.02$ & $12.37$ \\
    & Moving   & $75.20$ & $9.87$  & $10.96$ & $14.10$ \\
    & Noise    & $74.89$ & $10.23$  & $11.18$ & $12.58$ \\ \hline\hline
\multirow{3}{*}{MSIL}
    & \multicolumn{5}{l}{{\textit{Backbone Accuracy $A_{\text{baseline}}= 81.05\%$}}} \\
    & Static   & $80.72$ & $9.08$  & $8.53$  & $18.12$ \\
    & Moving   & $80.23$ & $12.03$ & $12.97$ & $14.06$ \\
    & Noise    & $81.02$ & $9.52$  & $8.68$  & $12.14$ \\ \hline\hline
\multirow{3}{*}{TE-Spikformer}
    & \multicolumn{5}{l}{{\textit{Backbone Accuracy $A_{\text{baseline}}= 80.92\%$}}} \\
    & Static   & $78.36$ & $9.07$  & $9.84$  & $18.08$ \\
    & Moving   & $79.91$ & $12.14$ & $9.50$  & $13.71$ \\
    & Noise    & $80.60$ & $9.58$  & $9.47$  & $12.12$ \\ 
\thickhline
\end{tabular}%
}
\end{table}

As demonstrated, our SpikeTimer mechanism consistently maintains high authorized accuracy ($A_{\text{auth}}$) and effectively suppresses unauthorized performance across four SNN architectures, Spikingformer, SD‑Transformer, MSIL, and TE‑Spikformer. Using the Static token type as a representative case, the authorized accuracy ranges from $74.63\%$ (SD‑Transformer) to $80.72\%$ (MSIL), with Spikingformer at $80.23\%$ and TE‑Spikformer at $78.36\%$. The $A_{\text{no}}$ ranges from $9.07\%$ (TE‑Spikformer) to $11.27\%$ (Spikingformer), and the false negative rate $A_{\text{mis}}$ ranges from $8.53\%$ (MSIL) to $11.02\%$ (SD‑Transformer). All unauthorized accuracies lie close to the random guessing baseline of $10\%$ for the 10‑class CIFAR10‑DVS task. Empirical results collectively confirm that SpikeTimer generalizes effectively across multiple SNN architectures and preserves its core functionality of selective temporal activation.

\subsection{Multi-User Capability}
\label{multi-user}

SpikeTimer ensures that models can differentiate between authorized and unauthorized inputs effectively. Building on this foundation, it is natural to explore the scalability of SpikeTimer beyond single-user scenarios, particularly in environments requiring multi-user access control.

We conducted the multi-user authorization experiment using static tokens on the N-MNIST and CIFAR10-DVS datasets.
This experiment extends the SpikeTimer framework to incorporate a multi-user authorization mechanism. Specifically, the authorized data is partitioned into subsets, each associated with a unique temporal slice representing different users (with the same token pattern). This design enables the model to authenticate multiple users while preserving the high performance of the original task. By embedding these user-specific tokens during training, the model learns to classify data based on both its inherent features and the temporal token, thereby supporting secure, user-specific access.

\begin{figure}[t]
    \centering
    \subcaptionbox{N-MNIST\protect\label{fig:5a}}
      {\includegraphics[width=0.49\linewidth]{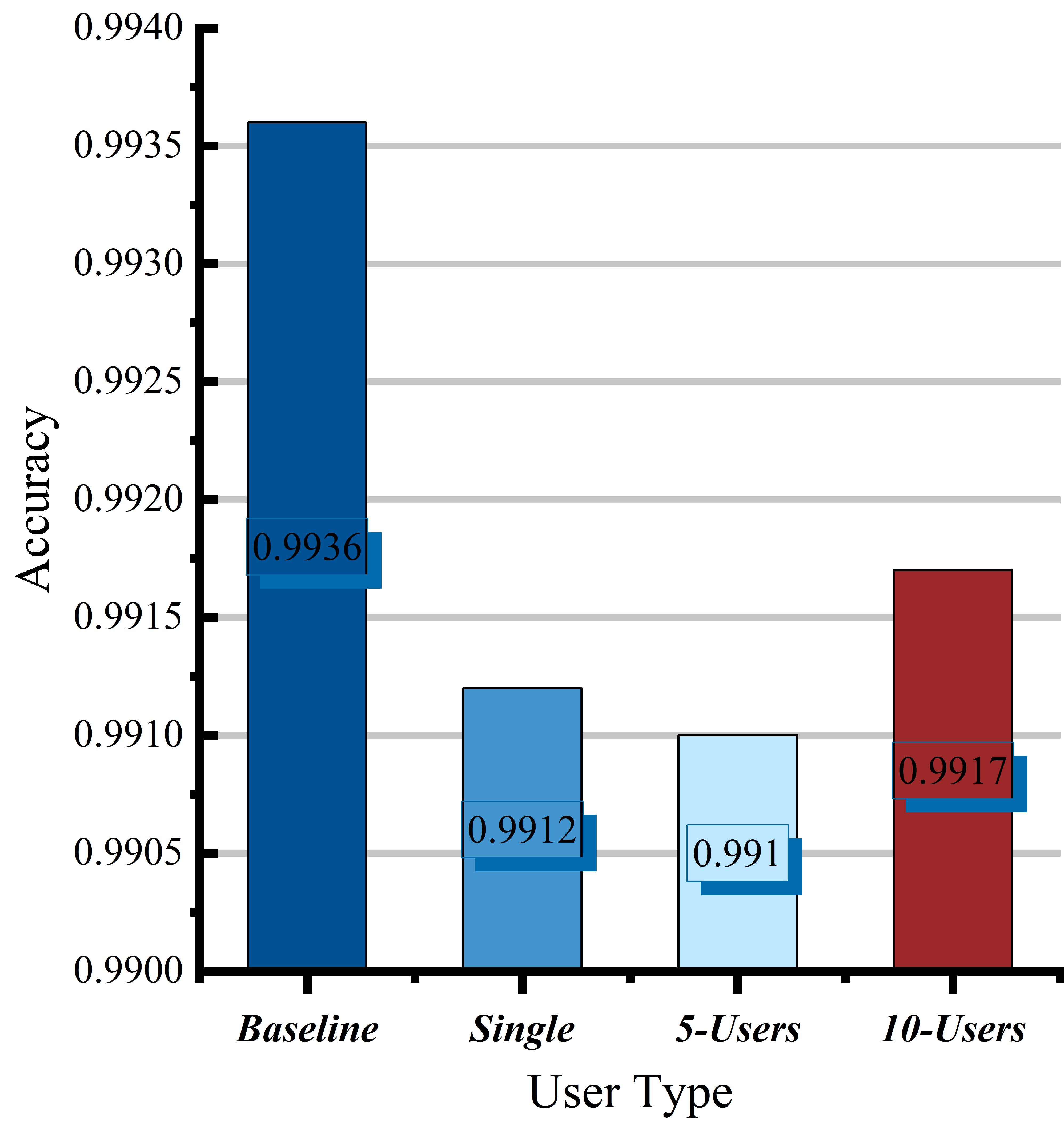}}
    \subcaptionbox{CIFAR10-DVS\protect\label{fig:5b}}
      {\includegraphics[width=0.49\linewidth]{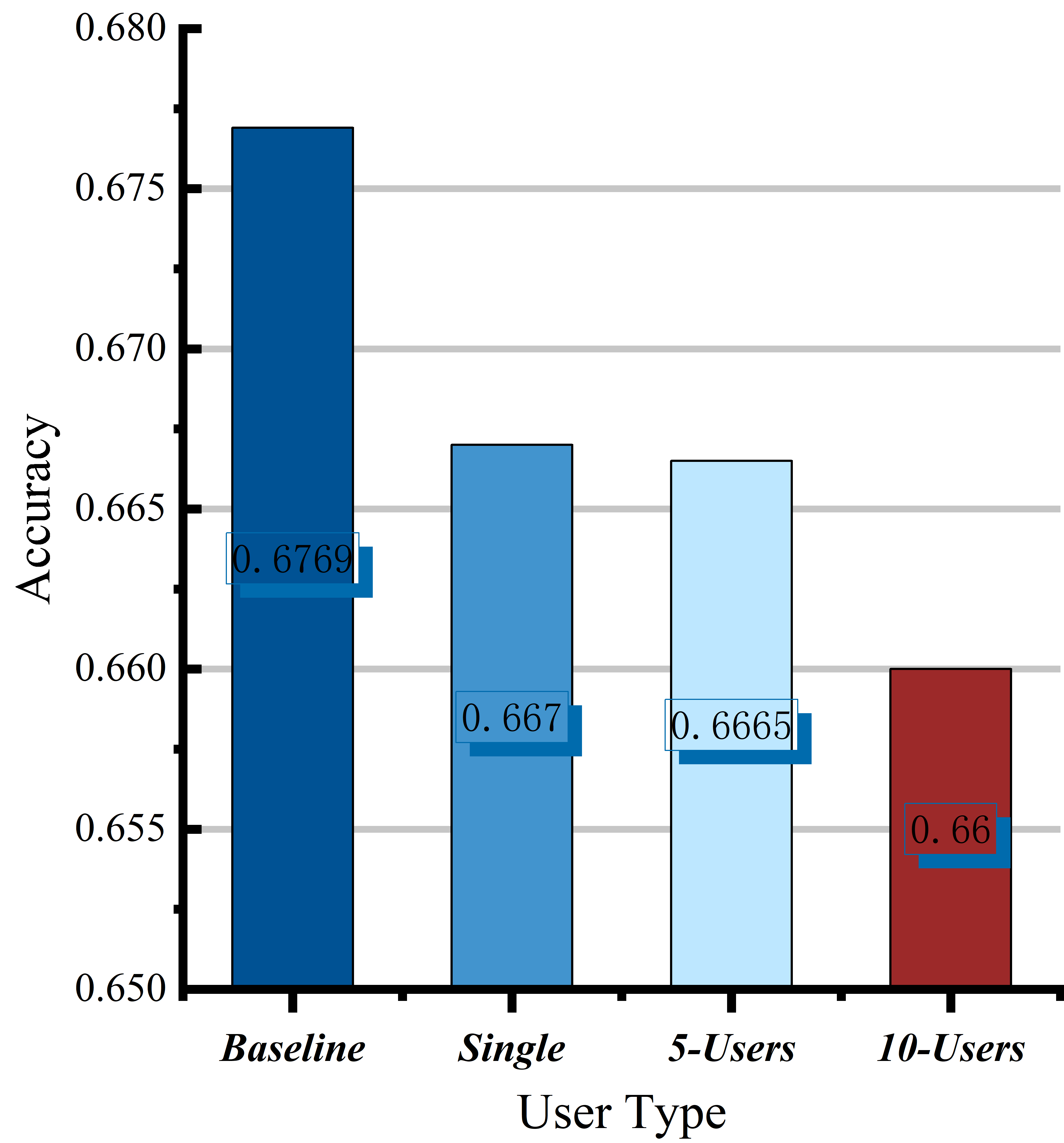}}
    \caption{Accuracy comparison for multi-user authorization on N-MNIST and CIFAR10-DVS. Baseline represents the performance of the model w/o adopting SpikeTimer.
    Please \textit{cf.} Sec.~\ref{multi-user} for detailed explanations.
    }
    \label{fig:5}
\end{figure}

The results are presented in Fig. \ref{fig:5}. On both datasets, it can be observed that in multi-user scenarios, the accuracy for each user approximately approaches the baseline accuracy. Notably, in multi-user scenarios, the accuracy for each user sometimes even slightly exceeds the accuracy observed in single-user scenarios. This may be explained by the implicit data augmentation effect introduced by user-specific tokens. 
The use of user-specific tokens diversifies the input spike trains for the SNN models. This introduces variability during training, akin to data augmentation, enabling the model to generalize better.
The results indicate that SpikeTimer can be extended to multi-user settings with negligible performance loss, proving its scalability for SNN copyright protection and user-specific adaptation.


\section{Conclusion}
\label{sect7}
This paper introduces SpikeTimer, a novel mechanism for securing SNNs through active protection via temporal token embedding. By leveraging the unique time-dependent nature of SNNs and neuromorphic data, SpikeTimer effectively mitigates unauthorized access, ensuring that only authorized tokens embedded within specific timeslices are able to activate the model. Through extensive experimentation, we validate that SpikeTimer achieves robust performance in both authorized and unauthorized scenarios, with marked suppression of no-token and mis-token accuracy. SpikeTimer can reduce unauthorized
data accuracy to random guess levels with only an
approximately $1.5\%$ drop in overall accuracy. 
Moreover, SpikeTimer substantiates strong generalization across multiple SNN architectures and natively supports multi-user authorization by assigning unique temporal tokens per user, granting exclusive model access through dedicated timeslices.
Furthermore, we demonstrate its resilience against various attacks, including fine-tuning and pruning.
For future work, we aim to investigate the performance of the method on large-scale, high-definition datasets, and cross-modal SNN protection.

\section{Acknowledgment}
This study is funded by the National Natural Science Foundation of China (Nos. 62572314 and 62471301). 
Yang Xiao and Gaolei Li contributed equally to this work and should be recognized as co-first authors.
Gaolei Li serves as the corresponding author.

\ifCLASSOPTIONcaptionsoff
  \newpage
\fi


\bibliographystyle{IEEEtran}
\bibliography{ref.bib}

@inproceedings{uchida2017embedding,
  title={Embedding watermarks into deep neural networks},
  author={Uchida, Yusuke and Nagai, Yuki and Sakazawa, Shigeyuki and Satoh, Shin'ichi},
  booktitle={Proc. ACM ICMR},
  pages={269--277},
  year={2017}
}

@inproceedings{deng2023surrogate,
  title={Surrogate module learning: Reduce the gradient error accumulation in training spiking neural networks},
  author={Deng, Shikuang and Lin, Hao and Li, Yuhang and Gu, Shi},
  booktitle={Proc. ICML},
  pages={7645--7657},
  year={2023}
}

@inproceedings{wang2021riga,
  title={Riga: Covert and robust white-box watermarking of deep neural networks},
  author={Wang, Tianhao and Kerschbaum, Florian},
  booktitle={Proc. WWW},
  pages={993--1004},
  year={2021}
}

@inproceedings{darvish2019deepsigns,
  title={Deepsigns: An end-to-end watermarking framework for ownership protection of deep neural networks},
  author={Darvish Rouhani, Bita and Chen, Huili and Koushanfar, Farinaz},
  booktitle={Proc. ASPLOS},
  pages={485--497},
  year={2019}
}

@inproceedings{adi2018turning,
  title={Turning your weakness into a strength: Watermarking deep neural networks by backdooring},
  author={Adi, Yossi and Baum, Carsten and Cisse, Moustapha and Pinkas, Benny and Keshet, Joseph},
  booktitle={Proc. USENIX Security},
  pages={1615--1631},
  year={2018}
}

@article{bi1998synaptic,
  title={Synaptic modifications in cultured hippocampal neurons: dependence on spike timing, synaptic strength, and postsynaptic cell type},
  author={Bi, Guo-qiang and Poo, Mu-ming},
  journal={J. Neurosci.},
  volume={18},
  number={24},
  pages={10464--10472},
  year={1998},
  publisher={Soc Neuroscience}
}

@inproceedings{fang2021incorporating,
  title={Incorporating learnable membrane time constant to enhance learning of spiking neural networks},
  author={Fang, Wei and Yu, Zhaofei and Chen, Yanqi and Masquelier, Timoth{\'e}e and Huang, Tiejun and Tian, Yonghong},
  booktitle={Proc. CVPR},
  pages={2661--2671},
  year={2021}
}

@inproceedings{liu2022spikeconverter,
  title={Spikeconverter: An efficient conversion framework zipping the gap between artificial neural networks and spiking neural networks},
  author={Liu, Fangxin and Zhao, Wenbo and Chen, Yongbiao and Wang, Zongwu and Jiang, Li},
  booktitle={Proc. AAAI},
  volume={36},
  number={2},
  pages={1692--1701},
  year={2022}
}

@inproceedings{kundu2021hire,
  title={Hire-snn: Harnessing the inherent robustness of energy-efficient deep spiking neural networks by training with crafted input noise},
  author={Kundu, Souvik and Pedram, Massoud and Beerel, Peter A},
  booktitle={Proc. CVPR},
  pages={5209--5218},
  year={2021}
}

@article{li2017cifar10,
  title={Cifar10-dvs: an event-stream dataset for object classification},
  author={Li, Hongmin and Liu, Hanchao and Ji, Xiangyang and Li, Guoqi and Shi, Luping},
  journal={Front. Neurosci.},
  volume={11},
  pages={309},
  year={2017},
  publisher={Frontiers Media SA}
}

@inproceedings{guo2019spottune,
  title={Spottune: transfer learning through adaptive fine-tuning},
  author={Guo, Yunhui and Shi, Honghui and Kumar, Abhishek and Grauman, Kristen and Rosing, Tajana and Feris, Rogerio},
  booktitle={Proc. CVPR},
  pages={4805--4814},
  year={2019}
}

@article{gao2020backdoor,
  title={Backdoor attacks and countermeasures on deep learning: A comprehensive review},
  author={Gao, Yansong and Doan, Bao Gia and Zhang, Zhi and Ma, Siqi and Zhang, Jiliang and Fu, Anmin and Nepal, Surya and Kim, Hyoungshick},
  journal={arXiv preprint arXiv:2007.10760},
  year={2020}
}

@article{guo2024neuroclip,
  title={Neuroclip: Neuromorphic data understanding by clip and snn},
  author={Guo, Yufei and Chen, Yuanpei and Ma, Zhe},
  journal={IEEE Signal Process. Lett.},
  year={2024},
  publisher={IEEE}
}

@inproceedings{wang2022signed,
  title={Signed Neuron with Memory: Towards Simple, Accurate and High-Efficient ANN-SNN Conversion.},
  author={Wang, Yuchen and Zhang, Malu and Chen, Yi and Qu, Hong},
  booktitle={Proc. IJCAI},
  pages={2501--2508},
  year={2022}
}

@article{liu2021discrimination,
  title={Discrimination-aware network pruning for deep model compression},
  author={Liu, Jing and Zhuang, Bohan and Zhuang, Zhuangwei and Guo, Yong and Huang, Junzhou and Zhu, Jinhui and Tan, Mingkui},
  journal={IEEE TPAMI},
  volume={44},
  number={8},
  pages={4035--4051},
  year={2021},
  publisher={IEEE}
}

@InProceedings{Cui_2024_CVPR,
    author    = {Cui, Qi and Meng, Ruohan and Xu, Chaohui and Chang, Chip-Hong},
    title     = {Steganographic Passport: An Owner and User Verifiable Credential for Deep Model IP Protection Without Retraining},
    booktitle = {Proc. CVPR},
    month     = {June},
    year      = {2024},
    pages     = {12302-12311}
}

@inproceedings{ren2024activedaemon,
  title={Active{D}aemon: Unconscious {DNN} Dormancy and Waking Up via User-specific Invisible Token},
  author={Ren, Ge and Li, Gaolei and Li, Shenghong and Chen, Libo and Ren, Kui},
  booktitle={Proc. NDSS},
  year={2024}
}

@article{ren2022protecting,
  title={Protecting intellectual property with reliable availability of learning models in ai-based cybersecurity services},
  author={Ren, Ge and Wu, Jun and Li, Gaolei and Li, Shenghong and Guizani, Mohsen},
  journal={IEEE TDSC},
  volume={21},
  number={2},
  pages={600--617},
  year={2022},
  publisher={IEEE}
}

@article{xue2022advparams,
  title={AdvParams: An active DNN intellectual property protection technique via adversarial perturbation based parameter encryption},
  author={Xue, Mingfu and Wu, Zhiyu and Zhang, Yushu and Wang, Jian and Liu, Weiqiang},
  journal={IEEE TETC},
  volume={11},
  number={3},
  pages={664--678},
  year={2022},
}

@article{lin2020chaotic,
  title={Chaotic weights: A novel approach to protect intellectual property of deep neural networks},
  author={Lin, Ning and Chen, Xiaoming and Lu, Hang and Li, Xiaowei},
  journal={IEEE TCAD},
  volume={40},
  number={7},
  pages={1327--1339},
  year={2020},
  publisher={IEEE}
}

@article{peng2023intellectual,
  title={Intellectual property protection of DNN models},
  author={Peng, Sen and Chen, Yufei and Xu, Jie and Chen, Zizhuo and Wang, Cong and Jia, Xiaohua},
  journal={World Wide Web},
  volume={26},
  number={4},
  pages={1877--1911},
  year={2023},
}

@inproceedings{amir2017low,
  title={A low power, fully event-based gesture recognition system},
  author={Amir, Arnon and others},
  booktitle={Proc. CVPR},
  pages={7243--7252},
  year={2017}
}

@article{orchard2015converting,
  title={Converting static image datasets to spiking neuromorphic datasets using saccades},
  author={Orchard, Garrick and Jayawant, Ajinkya and Cohen, Gregory K and Thakor, Nitish},
  journal={Front. Neurosci.},
  volume={9},
  pages={437},
  year={2015},
  publisher={Frontiers Media SA}
}

@article{eshraghian2023training,
  title={Training spiking neural networks using lessons from deep learning},
  author={Eshraghian, Jason K and Ward, Max and Neftci, Emre O and Wang, Xinxin and Lenz, Gregor and Dwivedi, Girish and Bennamoun, Mohammed and Jeong, Doo Seok and Lu, Wei D},
  journal={Proc. IEEE},
  year={2023},
  publisher={IEEE}
}

@inproceedings{cheng2020lisnn,
  title={LISNN: Improving spiking neural networks with lateral interactions for robust object recognition.},
  author={Cheng, Xiang and Hao, Yunzhe and Xu, Jiaming and Xu, Bo},
  booktitle={Proc. IJCAI},
  pages={1519--1525},
  year={2020},
}

@inproceedings{li2021free,
  title={A free lunch from ANN: Towards efficient, accurate spiking neural networks calibration},
  author={Li, Yuhang and Deng, Shikuang and Dong, Xin and Gong, Ruihao and Gu, Shi},
  booktitle={Proc. ICML},
  pages={6316--6325},
  year={2021}
}

@inproceedings{zhang2018protecting,
  title={Protecting intellectual property of deep neural networks with watermarking},
  author={Zhang, Jialong and Gu, Zhongshu and others},
  booktitle={Proc. ACM Asia CCS},
  pages={159--172},
  year={2018}
}

@inproceedings{han2020rmp,
  title={Rmp-snn: Residual membrane potential neuron for enabling deeper high-accuracy and low-latency spiking neural network},
  author={Han, Bing and Srinivasan, Gopalakrishnan and Roy, Kaushik},
  booktitle={Proc. CVPR},
  pages={13558--13567},
  year={2020}
}

@article{guo2023direct,
  title={Direct learning-based deep spiking neural networks: a review},
  author={Guo, Yufei and Huang, Xuhui and Ma, Zhe},
  journal={Front. Neurosci.},
  volume={17},
  pages={1209795},
  year={2023},
}

@inproceedings{jia2021entangled,
  title={Entangled watermarks as a defense against model extraction},
  author={Jia, Hengrui and Choquette-Choo, Christopher A and Chandrasekaran, Varun and Papernot, Nicolas},
  booktitle={Proc. USENIX Security},
  pages={1937--1954},
  year={2021}
}

@inproceedings{cong2022sslguard,
  title={Sslguard: A watermarking scheme for self-supervised learning pre-trained encoders},
  author={Cong, Tianshuo and He, Xinlei and Zhang, Yang},
  booktitle={Proc. ACM CCS},
  pages={579--593},
  year={2022}
}

@book{gerstner2002spiking,
  title={Spiking neuron models: Single neurons, populations, plasticity},
  author={Gerstner, Wulfram and Kistler, Werner M},
  year={2002},
  publisher={Cambridge university press}
}

@article{neftci2019surrogate,
  title={Surrogate gradient learning in spiking neural networks: Bringing the power of gradient-based optimization to spiking neural networks},
  author={Neftci, Emre O and Mostafa, Hesham and Zenke, Friedemann},
  journal={IEEE Signal Process. Mag.},
  volume={36},
  number={6},
  pages={51--63},
  year={2019},
  publisher={IEEE}
}

@inproceedings{rebecq2019events,
  title={Events-to-video: Bringing modern computer vision to event cameras},
  author={Rebecq, Henri and Ranftl, Ren{\'e} and Koltun, Vladlen and Scaramuzza, Davide},
  booktitle={Proc. CVPR},
  pages={3857--3866},
  year={2019}
}

@inproceedings{poursiami2024watermarking,
  title={Watermarking Neuromorphic Brains: Intellectual Property Protection in Spiking Neural Networks},
  author={Poursiami, Hamed and Alouani, Ihsen and Parsa, Maryam},
  booktitle={Proc. ICONS},
  pages={287--294},
  year={2024},
}

@inproceedings{DBLP:conf/aaai/ShanZ0QEQ25,
  author       = {Yimeng Shan and
                  Malu Zhang and
                  Ruijie Zhu and
                  Xuerui Qiu and
                  Jason K. Eshraghian and
                  Haicheng Qu},
  title        = {Advancing Spiking Neural Networks Towards Multiscale Spatiotemporal Interaction Learning},
  booktitle    = {Proc. AAAI},
  year         = {2025},
}

@article{
adi1480,
author = {Wei Fang  and Yanqi Chen  and Jianhao Ding  and Zhaofei Yu  and Timothée Masquelier  and Ding Chen  and Liwei Huang  and Huihui Zhou  and Guoqi Li  and Yonghong Tian },
title = {SpikingJelly: An open-source machine learning infrastructure platform for spike-based intelligence},
journal = {Sci. Adv.},
volume = {9},
number = {40},
pages = {eadi1480},
year = {2023},
}

@INPROCEEDINGS{8835365,
  author={Wang, Bolun and Yao, Yuanshun and Shan, Shawn and Li, Huiying and Viswanath, Bimal and Zheng, Haitao and Zhao, Ben Y.},
  booktitle={Proc. IEEE SP}, 
  title={Neural Cleanse: Identifying and Mitigating Backdoor Attacks in Neural Networks}, 
  year={2019},
  volume={},
  number={},
  pages={707-723},
}

@article{GAO2024128268,
title = {TE-Spikformer:Temporal-enhanced spiking neural network with transformer},
journal = {Neurocomputing},
volume = {602},
pages = {128268},
year = {2024},
author = {ShouWei Gao and XiangYu Fan and XingYang Deng and ZiChao Hong and Hao Zhou and ZiHao Zhu},
}

@inproceedings{10.1145/3746027.3755227,
author = {Yin, Yangxu and Chen, Honglong and Gao, Yudong and Sun, Peng and Wu, Liantao and Li, Zhe and Liu, Weifeng},
title = {FFCBA: Feature-based Full-target Clean-label Backdoor Attacks},
year = {2025},
booktitle = {Proc. ACM MM},
pages = {3884–3892},
numpages = {9},
}

@inproceedings{NEURIPS2024_4ce18228,
 author = {Xia, Jun and Yue, Zhihao and Zhou, Yingbo and Ling, Zhiwei and Shi, Yiyu and Wei, Xian and Chen, Mingsong},
 booktitle = {Proc. NeurIPS},
 pages = {43549-43570},
 publisher = {Curran Associates, Inc.},
 title = {WaveAttack: Asymmetric Frequency Obfuscation-based Backdoor Attacks Against Deep Neural Networks},
 year = {2024}
}

@inproceedings{DBLP:conf/aaai/ZhouYZZWZMT26,
  author       = {Chenlin Zhou and
                  Liutao Yu and
                  Zhaokun Zhou and
                  Han Zhang and
                  Jiaqi Wang and
                  Huihui Zhou and
                  Zhengyu Ma and
                  Yonghong Tian},
  title        = {Spikingformer: A Key Foundation Model for Spiking Neural Networks},
  booktitle    = {Proc. AAAI},
  pages        = {2236--2244},
  year         = {2026},
}

@inproceedings{DBLP:conf/eccv/WangXLYLZSBZ24,
  author       = {Qi Wang and
                  Zhou Xu and
                  Yuming Lin and
                  Jingtao Ye and
                  Hongsheng Li and
                  Guangming Zhu and
                  Syed Afaq Ali Shah and
                  Mohammed Bennamoun and
                  Liang Zhang},
  title        = {DailyDVS-200: {A} Comprehensive Benchmark Dataset for Event-Based
                  Action Recognition},
  booktitle    = {Proc. ECCV},
  pages        = {55--72},
  year         = {2024},
}

@ARTICLE{10.3389/fnins.2015.00437,
    
AUTHOR={Orchard, Garrick  and Jayawant, Ajinkya  and Cohen, Gregory K.  and Thakor, Nitish },
           
TITLE={Converting Static Image Datasets to Spiking Neuromorphic Datasets Using Saccades},
          
JOURNAL={Front. Neurosci.},
          
VOLUME={9},
  
YEAR={2015},
}

@inproceedings{NEURIPS2023_ca0f5358,
 author = {Yao, Man and Hu, JiaKui and Zhou, Zhaokun and Yuan, Li and Tian, Yonghong and Xu, Bo and Li, Guoqi},
 booktitle = {Proc. NeurIPS},
 pages = {64043--64058},
 title = {Spike-driven Transformer},
 year = {2023}
}

\begin{IEEEbiography}[{\includegraphics[width=1in,height=1.25in,clip,keepaspectratio]{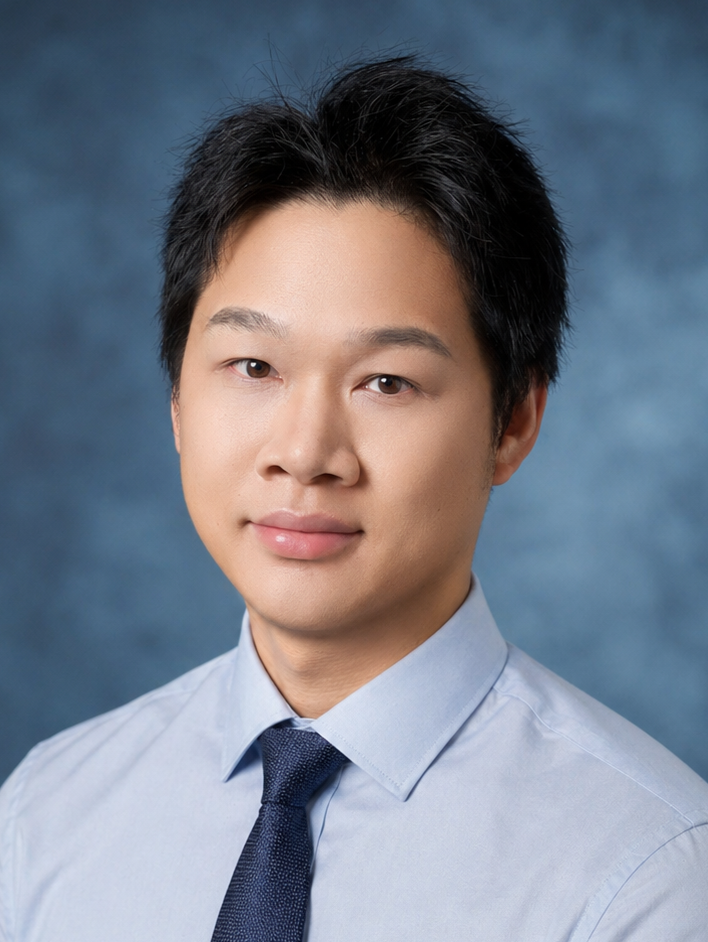}}]{Xiao Yang}(Member, IEEE) received his Ph.D. degree with distinction and M.S. degree in Computer Science from Shanghai Jiao Tong University, Shanghai, China. 
He previously served as Research Assistant with the Department of Computing, The Hong Kong Polytechnic University, Hung Hom, Kowloon, Hong Kong (2023–2024), and was a Ph.D. Research Student with the Department of Intelligence Science and Technology, Kyoto University, Kyoto, Japan (2021–2022).
He is the recipient of several Best Paper Awards at IEEE conferences.
His research interests center on trustworthy learning and network security.
\end{IEEEbiography}

\vspace{-10pt}
\begin{IEEEbiography}
[{\includegraphics[width=1in,height=1.25in,clip,keepaspectratio]{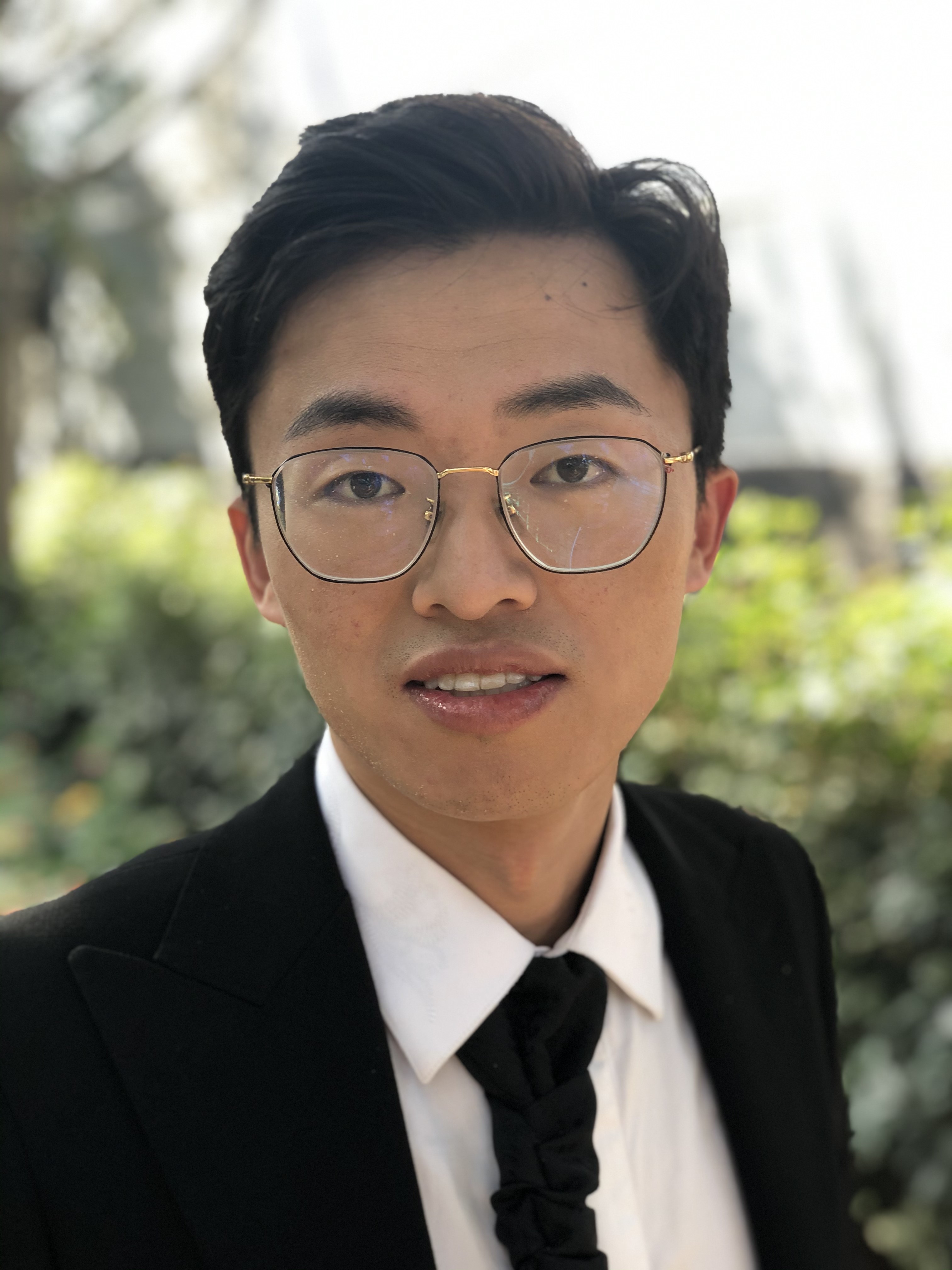}}]{Gaolei Li} (Member, IEEE) is currently an Associate Professor with the School of Computer Science, Shanghai Jiao Tong University, Shanghai, China. He received his B.S. degree from Sichuan University, Chengdu, China, and Ph.D. degree from Shanghai Jiao Tong University. From 2018 to 2019, he visited Muroran Institution of Technology, Hokkaido, Japan. 
He has published more than 120 papers on many top-tier conferences and journals like NDSS, ACM CCS, AAAI, IJCAI, ACM MM, IEEE TIFS, IEEE TDSC, IEEE TMC, etc. 
He has received the Best Paper Awards from IEEE ICNC, IEEE ISPA, IEEE ComSoc CSIM Committee, Chinese Association for Cryptologic Research, and Student Travel Grant Award for IEEE GLOBECOM.
His research interests include network security, communication security, and adversarial machine learning. 

\end{IEEEbiography}
\vspace{-10pt}
\begin{IEEEbiography}[{\includegraphics[width=1in,height=1.2in,clip,keepaspectratio]{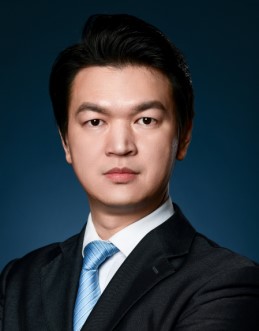}}]{Jun Wu} (Senior Member, IEEE) is currently a Professor and the Associate Dean in School of Computer Science, Shanghai Jiao Tong University, Shanghai, China. He is also the Dean of Institute of Cyber Science and Technology and Associate Director of National Engineering Research Center for Information Content Analysis Technology, Shanghai Jiao Tong University.
He received the Ph.D. degree from Waseda University, Tokyo, Japan, in 2011. 
He is the Chair of IEEE P21451-1-5 Standard Working Group. 
He is an Associate Editor of IEEE Systems Journal and IEEE Networking Letters. He was a Guest Editor of IEEE Transactions on Industrial Informatics, IEEE Transactions on Intelligent Transportation Systems, IEEE Sensors Journal, Sensors, and Frontiers of Information Technology and Electronic Engineering.
His research interests include the intelligence and security techniques of artificial intelligence, Internet of Things (IoT), smart network, molecular communication, etc. 
\end{IEEEbiography}
\vspace{-10pt}
\begin{IEEEbiography}
[{\includegraphics[width=1in,height=1.2in,clip,keepaspectratio]{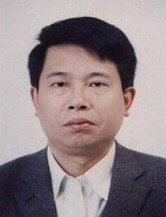}}]{Jianhua Li} (Senior Member, IEEE) is a Professor and the Dean of the Institute of Cyber Science and Technology, Shanghai Jiao Tong University, Shanghai, China. He is also the Director of National Engineering Laboratory for Information Content Analysis Technology and the Director of Engineering Research Center for Network Information Security Management and Service of Chinese Ministry of Education. 
He has published more than 400 papers, including IEEE TIFS, IEEE TDSC, IEEE TMC, IEEE TKDE, IEEE INFOCOM, ACM CCS, ACM MobiHoc, CVPR, etc. He published 6 books and has about 20 patents. He made 3 standards and has 5 software copyrights. 
He has received many awards, including the Second Prize of National Technology Progress Award of China in 2005, the first Prize of Science and Technology Progress Award of the Ministry of Education, the first Prize of Science and Technology Progress Award of Shanghai, IEEE CSIM Committee Best Paper Award in 2016, IEEE TETC Best Paper Award in 2017, and ESI Highly Cited Scientist from 2022 to 2024.
His work has been recognized with numerous awards, including the Second Prize of National Technology Progress Award of China, the first Prize of Science and Technology Progress Award of the Ministry of Education, and the first Prize of Science and Technology Progress Award of Shanghai.
His research interests center on information security.
\end{IEEEbiography}

\vspace{-10pt}
\begin{IEEEbiography}[{\includegraphics[width=1in,height=1.2in,clip,keepaspectratio]{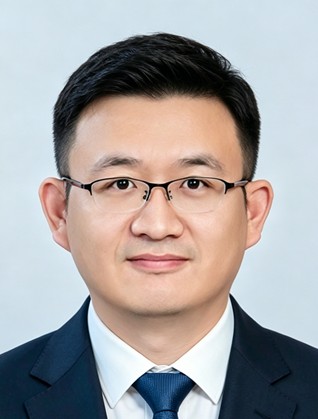}}]{Zhiquan Liu} (Senior Member, IEEE) received the B.S. degree from the School of Science, Xidian University, Xi'an, China, in 2012, and the Ph.D. degree from the School of Computer Science and Technology, Xidian University in 2017. He is currently the Associate Dean and a Full Professor at the College of Cyber Security, Jinan University, Guangzhou, China. His current research focuses on security, trust, privacy, and intelligence in vehicular networks. He currently serves as the Area Editor or Associate Editor of multiple SCI-index journals, such as IEEE TIFS, IEEE TDSC, IEEE TKDE, IEEE TPAMI, IEEE TCSVT, IEEE TII, Information Fusion, Computer Networks, Ad Hoc Networks, etc. 
\end{IEEEbiography}

\end{document}